\documentclass[pre,twocolumn,preprintnumbers,amsmath,amssymb,nofootinbib,floatfix]{revtex4}

\usepackage{graphicx,bm,float}

\makeatletter
\def\graphicscale{\twocolumn@sw{0.3}{0.4}}
\def\graphicthreescale{\twocolumn@sw{0.3}{0.4}}

\begin{document}

\title{Multicomponent compact Abelian-Higgs lattice models}

\author{Andrea Pelissetto}
\affiliation{Dipartimento di Fisica dell'Universit\`a di Roma Sapienza
        and INFN Sezione di Roma I, I-00185 Roma, Italy}

\author{Ettore Vicari} 
\affiliation{Dipartimento di Fisica dell'Universit\`a di Pisa
        and INFN Largo Pontecorvo 3, I-56127 Pisa, Italy}

\date{\today}

\begin{abstract}
We investigate the phase diagram and critical behavior of
three-dimensional multicomponent Abelian-Higgs models, in which an
$N$-component complex field $z^a_{\bm x}$ of unit length and charge is
coupled to compact quantum electrodynamics in the usual Wilson lattice
formulation.  We determine the phase diagram and study the nature of
the transition line for $N=2$ and $N=4$.  Two phases are identified,
specified by the behavior of the gauge-invariant local composite
operator $Q_{\bm x}^{ab} = \bar{z}^a_{\bm x} z^b_{\bm x} -
\delta^{ab}/N$, which plays the role of order parameter.  In one
phase, we have $\langle Q_{\bm x}^{ab}\rangle =0$, while in the other
$Q_{\bm x}^{ab}$ condenses. Gauge correlations are never critical:
gauge excitations are massive for any finite coupling. The two phases
are separated by a transition line. Our numerical data are consistent
with the simple scenario in which the nature of the transition is
independent of the gauge coupling.  Therefore, for any finite positive
value of the gauge coupling, we predict a continuous transition in the
Heisenberg universality class for $N=2$ and a first-order transition
for $N=4$. However, notable crossover phenomena emerge for large gauge
couplings, when gauge fluctuations are suppressed.  Such crossover
phenomena are related to the unstable O($2N$) fixed point, describing
the behavior of the model in the infinite gauge-coupling limit.
\end{abstract}

\maketitle


\section{Introduction}
\label{intro}

Models of complex scalar matter fields coupled to gauge fields have
been much studied in condensed matter physics, since they are believed
to describe several interesting systems, such as superconductors and
superfluids, quantum Hall states, quantum SU($N$) antiferromagnets,
unconventional quantum phase transitions, etc., see, {\em e.g.},
Refs.~\cite{SSNHS-03,SSS-04,BSA-04,MV-04,SBSVF-04,KHI-11} and
references therein.  Scalar electrodynamics, or Abelian-Higgs (AH)
model, is a paradigmatic model, in which a $N$-component complex
scalar field ${\bm \Phi}$ is minimally coupled to the electromagnetic
field $A_\mu$. The corresponding continuum Lagrangian reads
\begin{equation}
{\cal L} = 
|D_\mu{\bm\Phi}|^2  + r\, |{\bm \Phi}|^2 + 
{1\over 6} u \,(|{\bm \Phi}|^2)^2
+ {1\over 4 g^2} \,F_{\mu\nu}^2 
\,,
\label{abhim}
\end{equation}
where $F_{\mu\nu}\equiv \partial_\mu A_\nu - \partial_\nu A_\mu$, and
$D_\mu \equiv \partial_\mu + i A_\mu$.  The renormalization-group (RG)
analysis of the continuum AH model~\cite{HLM-74,MZ-03} should provide
information on the nature of the finite-temperature phase transitions
occurring in $d$-dimensional systems characterized by a global U($N$)
symmetry and a local U(1) gauge symmetry.

In this paper we consider the multicomponent AH model, in which the
scalar field ${\bm\Phi}$ has $N\ge 2$ components. Such a model has a
local U(1) gauge invariance and a global U($N$) invariance. We assume
that the field belongs to the fundamental representation of the U(1)
group, i.e., it has charge 1. The one-component AH model has been
extensively discussed in the literature
\cite{KKLP-98,KNS-02,MHS-02,WBJSS-05,CIS-06}. In three dimensions,
these systems may undergo continuous transitions in the XY
universality class.

Lattice formulations of the three-dimensional AH model are obtained by
associating complex $N$-component unit vectors ${\bm z}_{\bm x}$ with
the sites ${\bm x}$ of a cubic lattice, and U(1) variables
$\lambda_{{\bm x},\mu}$ with each link connecting the site ${\bm x}$
with ${\bm x}+\hat\mu$ (where $\hat\mu=\hat{1},\hat{2},\ldots$ are
unit vectors along the lattice directions).  The partition function of
the system reads
\begin{equation}
Z = \sum_{\{{\bm z}\},\{\lambda\}} e^{-H}\,,
\label{partfun}
\end{equation}
where the Hamiltonian is~\cite{footnoteH}
\begin{eqnarray}
H &=& - \beta N
\sum_{{\bm x}, \mu}
\left( \bar{\bm{z}}_{\bm x} 
\cdot \lambda_{{\bm x},\mu}\, {\bm z}_{{\bm x}+\hat\mu} 
+ {\rm c.c.}\right)
\label{gllf}\\
&-& \beta_g \sum_{{\bm x},\mu\neq\nu} 
\left(
\lambda_{{\bm x},{\mu}} \,\lambda_{{\bm x}+\hat{\mu},{\nu}} 
\,\bar{\lambda}_{{\bm x}+\hat{\nu},{\mu}}  
  \,\bar{\lambda}_{{\bm x},{\nu}} 
+ {\rm c.c.}\right)\,,
\nonumber
\end{eqnarray}
where the first sum is over all links of the lattice, while the second
one is over all plaquettes.  

For $\beta_g=0$ we recover a particular lattice formulation of the
CP$^{N-1}$ model, which is quadratic with respect to the spin
variables and contains explicit gauge link variables. The CP$^{N-1}$
model has been extensively studied.  In spite of several
field-theoretical and numerical studies for $N=2,3,4$ and
$N\to\infty$, there are still some controversies on the nature of its
transition \cite{MZ-03,KHI-11,NCSOS-11,NCSOS-13,DPV-15,PTV-17,PV-19}.
For $\beta_g\to\infty$ the gauge link variables are all equal to 1
modulo gauge transformations and the AH model becomes equivalent to
the standard O($n$) vector model with $n=2N$, whose critical behavior
is well understood \cite{PV-02}.  We also mention that some numerical
results for the AH lattice model (\ref{gllf}) have been reported in
Refs.~\cite{KHI-11,ODHIM-07,TIM-05}, but a definite picture has not
been achieved yet~\cite{footnoteold}.

It is important to stress that we consider the lattice compact version
of electrodynamics (the so-called Wilson lattice formulation of gauge
theories).  In the absence of matter fields, its behavior \cite{QED}
is controlled by topological excitations, the monopoles, which are
instead suppressed in noncompact formulations. Therefore, the critical
properties of the AH lattice model that we consider might differ from
those of the model in which gauge fields are noncompact.

In this paper we investigate the phase diagram and the nature of the
phase transitions of the three-dimensional AH model (\ref{gllf}).  We
consider systems with $N=2$ and $N=4$, and investigate the nature of
the transition line by varying $\beta$ at fixed gauge coupling
$\beta_g$, for some values of $\beta_g$.  In both cases the phase
diagram of the AH lattice model (\ref{gllf}) turns out to present two
phases: for small $\beta$ there is a disordered confined phase, while
for large values of $\beta$ there is an ordered Higgs phase in which
correlations of the gauge-invariant hermitian operator $Q_{\bm x}^{ab}
= \bar{z}_{\bm x}^a {z}_{\bm x}^b - \delta^{ab}/N$ show long-range
order.  In both phases, and also along the transition line, the
correlations of gauge variables do not show critical behaviors. The
gauge coupling $\beta_g$ does not play any significant role: the
features of two phases are the same for any finite $\beta_g$.  The two
phases are separated by a single transition line, which connects the
CP$^{N-1}$ transition point ($\beta_g = 0$) to the O($2N$) transition
point ($\beta_g = \infty$) in the space of the two parameters $\beta$
and $\beta_g$.  For $\beta_g=0$ the transition is continuous for $N=2$
(belonging to the Heisenberg universality class) and of first order
for $N=4$.  We conjecture that the nature of the transitions along the
line separating the Higgs and confined phases does not change with
$\beta_g$.  Therefore, the transition is always continuous
(discontinuous) for $N=2$ ($N=4$).  We also observe significant
deviations for $\beta_g$ large ($\beta_g\gtrsim 1$), i.e., when gauge
fluctuations are suppressed. They are interpreted as a crossover
phenomenon due to the presence of a O($2N$) vector transition in the
limit $\beta_g\to\infty$.

The paper is organized as follows. In Sec.~\ref{FTanalysis} we review
the field-theoretical results for the AH model. In Sec.~\ref{rgeps} we
review the $\varepsilon$-expansion predictions obtained in the
continuum AH model, and in Sec.~\ref{GLW} we present instead the
results of the Landau-Ginzburg-Wilson (LGW) approach based on a
gauge-invariant order parameter. The two approaches are critically
compared in Sec.~\ref{compFT}.  The numerical results are presented in
Sec.~\ref{numres}.  The definitions of the quantities we consider are
given in Sec.~\ref{nusim}, while Sec.~\ref{N2res} and \ref{N4res}
present our results for $N=2$ and 4, respectively, focusing on the
behavior of the gauge-invariant order parameter.  Results for vector
and gauge observables are presented in Sec.~\ref{Vector}.  In
Sec.~\ref{Conclusions} we summarize and present our conclusions.  In
App.~\ref{FSSON} we discuss the limit $\beta_g\to\infty$. More details
on the behavior of the different observables in this limit are given
in the supplementary material \cite{suppl-mat}.

\section{Field theoretical approaches}
\label{FTanalysis}

In this section we outline some apparently alternative
field-theoretical approaches which can be employed to infer the nature
of the phase transitions in systems characterized by a U($N$) global
symmetry and a local U(1) gauge symmetry, such as the AH lattice
model.

\subsection{Renormalization-group flow 
in the AH model close to four dimensions}
\label{rgeps}

We now summarize the main features of the RG flow in the continuum AH
model (\ref{abhim}), which has been analyzed close to four dimensions
in the $\varepsilon\equiv 4-d$ expansion
framework~\cite{HLM-74,FH-96,IZMHS-19}, and in the large-$N$
limit~\cite{MZ-03}.

Close to four dimensions, the RG flow in the space of the renormalized
couplings $u$ and $f\equiv g^2$ (we rescale them as $u\to u/(24\pi^2)$
and $f\to f/(24\pi^2)$ to simplify the equations) can be computed in
perturbation theory. At one loop, the $\beta$ functions
read~\cite{HLM-74}
\begin{eqnarray}
&&\beta_u \equiv \mu {\partial u\over \partial \mu}
= - \varepsilon u + 
(N+4) u^2 - 18 u f + 54 f^2\,,
\nonumber \\
&&\beta_f \equiv \mu {\partial f \over \partial \mu}
= - \varepsilon f + N f^2\,.
\label{betafunc}
\end{eqnarray} 
One can easily verify that a stable fixed point exists only for $N >
N_c(\varepsilon)$,  with 
\begin{equation}
N_c(\varepsilon) =  N_{4}
+ O(\varepsilon)\,,\quad N_{4} = 90 +
24\sqrt{15} \approx 183\,.
\label{nceps}
\end{equation}
The corresponding zero of the $\beta$ functions is 
\begin{eqnarray}
&&f^* = {\varepsilon \over N}\,,\label{fstar}\\
&&u^*  = 
{N+18 + \sqrt{N^2-180N-540}\over 2 N(N+4)}\,\varepsilon
\approx {\varepsilon\over N}\,.
\label{ufstar}
\end{eqnarray}
The presence of a stable fixed point indicates that these systems may
undergo a continuous transition if $N$ is large enough [$N>N_c(1)$ in
  three dimensions], in agreement with the direct large-$N$
analysis~\cite{MZ-03}.  The qualitative picture obtained in the
one-loop calculation is not changed by higher-order calculations.  The
perturbative expansion has been recently extended to four loops
\cite{IZMHS-19}, obtaining $N_c(\varepsilon)$ to $O(\varepsilon^3)$,
\begin{equation}
N_c(\varepsilon) = N_{4}\left[1 - 1.752 \,\varepsilon + 0.789\,
\varepsilon^2 + 0.362 \,\varepsilon^3+ O(\varepsilon^4)\right] .
\end{equation}
The large coefficients make a reliable three-dimensional
($\varepsilon=1$) estimate quite problematic.  Nevertheless, by means
of a resummation of the expansion, Ref.~\cite{IZMHS-19} obtained the
estimate $N_c = 12.2(3.9)$ in three dimensions, which confirms the
absence of a stable fixed point for small values of $N$.

In the limit $\beta_g \to\infty$, the lattice AH model (\ref{gllf}) is
equivalent to the symmetric O$(2N)$ vector theory. Therefore, for
large $\beta_g$ one expects significant crossover effects, which
increase as $\beta_g$ increases, due to the nearby O$(2N)$ critical
behavior.  In the continuum AH model, the crossover is controlled by
the RG flow in the vicinity of the O$(2N)$ fixed point
\begin{equation}
u^*_{{\rm O}(2N)}= {1\over N+4}  \varepsilon\,,\qquad
f=0\,.
\label{O2nfp}
\end{equation}
This fixed point exists for any $N$ and is always unstable.  The
analysis of the stability matrix $\Omega_{ij} = \partial
\beta_i/\partial g_j$ shows that it has a positive eigenvalue
$\lambda_u=\omega$, where $\omega>0$ is the exponent controlling the
leading scaling corrections in O$(2N)$ vector models~\cite{PV-02}, and
a negative eigenvalue, which gives the dimension of the operator that
controls the crossover behavior,
\begin{equation}
\lambda_f = \left. {\partial \beta_f \over \partial f} \right|_{f=0,u=u^*}
 \; .
\end{equation}
Since the $\beta$-function $\beta_f(u,f)$ 
associated with $f$ has the general form
\begin{equation}
\beta_f = -\varepsilon f + f^2 F(u,f)
\label{betafform}
\end{equation}
 where $F(u,f)$ has a regular perturbative expansion (see, e.g., the
 four-loop expansion reported in Ref.~\cite{IZMHS-19}), we obtain
\begin{equation}
\lambda_f = -\varepsilon
\label{lambdafest}
\end{equation}
to all orders in perturbation theory. Therefore, the crossover exponent 
$y_f=-\lambda_f$ is 1 in
three dimensions.  Note that these crossover features related to the
unstable O($2N$) fixed point are independent of the existence of the
stable fixed point, which is only relevant to predict the eventual
asymptotic behavior.

\subsection{Gauge-invariant Landau-Ginzburg-Wilson framework}
\label{GLW}

An alternative field-theoretical approach is the LGW
framework~\cite{Landau-book,WK-74,Fisher-75,Ma-book,PV-02,PV-19}, in
which one assumes that the relevant critical modes are associated with
the gauge-invariant local composite site variable
\begin{equation}
Q_{{\bm x}}^{ab} = \bar{z}_{\bm x}^a z_{\bm x}^b - {1\over N}
\delta^{ab},
\label{qdef}
\end{equation}
which is a hermitian and traceless $N\times N$ matrix.  As discussed
in Refs.~\cite{PTV-17,PTV-18,PV-19}, this is a highly nontrivial
assumption, as it postulates that gauge
fields do not play an important role in the effective theory.  
The order-parameter field in
the corresponding LGW theory is therefore a traceless hermitian matrix
field $\Psi^{ab}({\bm x})$, which can be formally defined as the
average of $Q_{\bm x}^{ab}$ over a large but finite lattice domain.
The LGW field theory is obtained by considering the most general
fourth-order polynomial in $\Psi$ consistent with the U($N$) global
symmetry:
\begin{eqnarray}
{\cal H}_{\rm LGW} &=& {\rm Tr} (\partial_\mu \Psi)^2 
+ r \,{\rm Tr} \,\Psi^2 \label{hlg}\\
&+&   w \,{\rm tr} \,\Psi^3 
+  \,u\, ({\rm Tr} \,\Psi^2)^2  + v\, {\rm Tr}\, \Psi^4 .
\nonumber
\end{eqnarray}
Also in this framework continuous transitions may only arise if the RG
flow in the LGW theory has a stable fixed point.

For $N=2$, the cubic term in Eq.~(\ref{hlg}) vanishes and the two
quartic terms are equivalent.  Therefore, one recovers the
O(3)-symmetric vector LGW theory, leading to the prediction that the
phase transition may be continuous and, in this case, that it belongs to
the Heisenberg universality class. For $N\ge 3$, the cubic term is
generally expected to be present.  Its presence is usually taken as an
indication that phase transitions occurring in this class of systems
are generally of first order.  Indeed, a straightforward mean-field
analysis shows that the transition is of first order in four
dimensions where mean field applies. If statistical fluctuations are
small---this is the basic assumption---the transition should be of
first order also in three dimensions.  In this scenario, continuous
transitions may still occur, but they require a fine tuning of the
microscopic parameters leading to the effective cancellation of the
cubic term.  These arguments were originally \cite{DPV-15,PV-19}
applied to predict the behavior of CP$^{N-1}$ models. However, as they
are only based on symmetry considerations, they can be extended to AH
lattice models, as well.

\subsection{Comparison of the alternative field-theoretical approaches}
\label{compFT}

The two field-theoretical approaches outlined above give inconsistent
predictions both for small and large values of $N$.  The contradiction
is quite striking for the two-component $N=2$ case.  For this value of
$N$, the continuum AH model predicts the absence of continuous
transitions, due to the absence of a stable fixed point. On the other
hand, a stable fixed point---it is the usual Heisenberg O(3) fixed
point---exists in the effective LGW theory based on a gauge-invariant
order parameter, leaving open the possibility of observing continuous
transitions (first-order transitions are never excluded as the
statistical model may be outside the attraction domain of the fixed
point). The numerical results for the CP$^1$ lattice
models~\cite{NCSOS-11,PV-19}, as well as the AH lattice results we
shall present below, confirm the existence of continuous transitions
in models with $N=2$: the LGW theory provides the correct description
of the large-scale behavior of these systems. There are at least two
possible explanations for the failure of the continuum AH model. A
first possibility is that it does not encode the relevant degrees of
freedom at the transition. A second possibility is that the problem is
not in the continuum AH model, but rather in the perturbative
treatment around four dimensions. The three-dimensional fixed point
may not be analytically related to a four-dimensional fixed point, and
therefore it escapes any perturbative analysis in powers of
$\varepsilon$.

We also recall that the perturbative AH approach of Sec.~\ref{rgeps}
also fails for $N=1$. Although no stable fixed point is identified in
the $\varepsilon$ expansion, see Sec.~\ref{rgeps}, these models may
undergo continuous transitions in the XY universality class
\cite{KKLP-98,KNS-02,MHS-02}.  It is worth mentioning that there are
also other systems in which the $\varepsilon$ expansion fails to
provide the correct physical picture in three dimensions. We mention
the $\phi^4$ theories describing frustrated spin models with
noncollinear order~\cite{CPPV-04,NO-14} and the $^3$He superfluid
transition from the normal to the planar phase~\cite{DPV-04}.

For large values of $N$, the continuum AH theory and the effective LGW
approach give again contradictory results. Indeed, the former approach
indicates that continuous transitions are possible, a prediction which
is supported by the large-$N$ analysis of lattice models, see, e.g.,
Ref.~\cite{PV-19}.  If one trusts the argument based on the relevance
of the cubic term, the LGW approach predicts instead a first-order
transition, unless a fine tuning of the microscopic parameters is
performed.  Again, there are two possible explanations for the
different conclusions obtained in the LGW approach.  A first
possibility is that the critical modes at the transition are not
exclusively associated with the gauge-invariant order parameter $Q$
defined in Eq.~(\ref{qdef}). Other features, for instance the gauge
degrees of freedom, may become relevant, requiring an effective
description different from that of the LGW theory (\ref{hlg}). If this
interpretation is correct, the continuum AH model would be the correct
theory as it includes the gauge fields explicitly.  A second
possibility is that the presence of a cubic term in the LGW
Hamiltonian does not necessarily imply the absence of continuous
transitions in three dimensions, as it is usually assumed. It might be
that statistical fluctuations soften the transition as one moves from
four to three dimensions; see, e.g., Refs.~\cite{NCSOS-11,NCSOS-13} for
a discussion of this issue.

While the two field-theoretical approaches give different predictions
for $N=1,2$ and $N$ large (more precisely, for $N>N_c$, see
Sec.~\ref{rgeps}), for $3 \le N < N_c$ they both predict that all
models undergo a first-order transition. For $N=3$ simulation results
do not presently confirm it.  Indeed, while numerical results for the
lattice model with Hamiltonian (\ref{gllf}) and $\beta_g = 0$ show a
robust indication that the transition is of first order~\cite{PV-19},
the results for the loop model considered in
Refs.~\cite{NCSOS-11,NCSOS-13} apparently favor a continuous
transition.  The available numerical results for lattice CP$^3$
models, i.e., for $N=4$, are generally consistent with first-order
transitions~\cite{KHI-11,NCSOS-13,PV-19}.  We also mention that
Ref.~\cite{KHI-11} claims that the AH lattice model (\ref{gllf})
undergoes a continuous transition for $\beta_g=1$ and $N=4$, a result
which is at odds with the above arguments. However, as we shall show,
the numerical results that we present later do not confirm their
conclusions, but are instead consistent with a relatively weak
first-order transition.

\section{Numerical results}
\label{numres}

\subsection{Numerical simulations and observables}
\label{nusim}

\begin{figure}[tbp]
\includegraphics*[scale=\graphicscale]{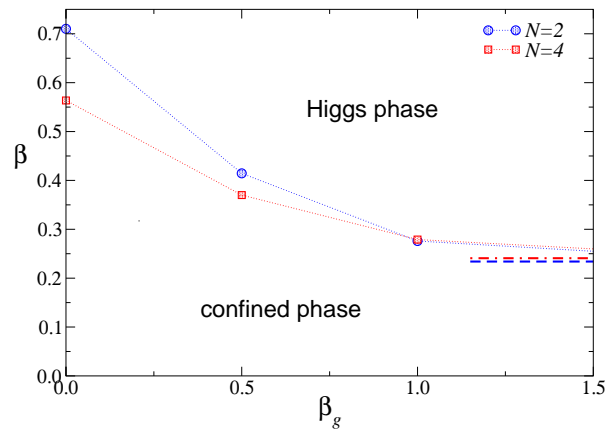}
\caption{ The phase diagram of the AH lattice model in the space of
  parameters $\beta$ and $\beta_g$, for $N=2$ and $N=4$. The points
  are the MC estimates of the critical points, the dotted lines that
  connect them are only meant to guide the eye.  The horizontal lines
  indicate the limiting values of $\beta_c$ for $\beta_g\to \infty$
  for $N=2$ ($\beta_c\approx 0.23396$, dashed) and $N=4$
  ($\beta_c\approx 0.24084$, dot-dashed): they correspond to the
  critical $\beta_c$ for the standard three-dimensional O$(4)$ and
  O$(8)$ vector models, respectively. For $\beta_g = 0$, we have \cite{PV-19}
  $\beta_c = 0.7102(1)$ ($N=2$) and $\beta_c = 0.5636(1)$ ($N=4$).
}
\label{betac}
\end{figure}

In this section we present a finite-size scaling (FSS) analysis of
numerical results of Monte Carlo (MC) simulations for $N=2$ and $N=4$.
For this purpose we consider cubic lattices of linear size $L$ with
periodic boundary conditions. We study the behavior of the system as a
function of $\beta$ at fixed $\beta_g$.

The linearity of Hamiltonian (\ref{gllf}) with respect to each lattice
variable allows us to employ an overrelaxed algorithm for the updating
of the lattice configurations.  It consists in a stochastic mixing of
microcanonical and standard Metropolis updates of the lattice
variables~\cite{CRV-92,DMV-04,Hasenbusch-17}.  To update each lattice
variable, we randomly choose either a standard Metropolis update,
which ensures ergodicity, or a microcanonical move, which is more
efficient than the Metropolis one but does not change the energy. On
average, we perform three/four microcanonical updates for every
Metropolis proposal.  In the Metropolis update, changes are tuned so
that the acceptance is 1/3.

We compute the energy density and the specific heat, defined as
\begin{eqnarray}
E = {1\over N V} \langle H \rangle,\qquad
C ={1\over N^2 V}
\left( \langle H^2 \rangle 
- \langle H  \rangle^2\right),
\label{ecvdef}
\end{eqnarray}
where $V=L^3$.  We consider correlations of the hermitean gauge
invariant operator (\ref{qdef}). Its two-point correlation function is
defined as
\begin{equation}
G({\bm x}-{\bm y}) = \langle {\rm Tr}\, Q_{\bm x} Q_{\bm y} \rangle,  
\label{gxyp}
\end{equation}
where the translation invariance of the system has been taken into
account.  The susceptibility and the correlation length are defined as
$\chi=\sum_{\bm x} G({\bm x})$ and
\begin{eqnarray}
\xi^2 \equiv  {1\over 4 \sin^2 (\pi/L)}
{\widetilde{G}({\bm 0}) - \widetilde{G}({\bm p}_m)\over 
\widetilde{G}({\bm p}_m)},
\label{xidefpb}
\end{eqnarray}
where $\widetilde{G}({\bm p})=\sum_{{\bm x}} e^{i{\bm p}\cdot {\bm x}}
G({\bm x})$ is the Fourier transform of $G({\bm x})$, and ${\bm p}_m =
(2\pi/L,0,0)$ is the minimum nonzero lattice momentum.  We also
consider the Binder parameter
\begin{equation}
U = {\langle \mu_2^2\rangle \over \langle \mu_2 \rangle^2} \,, \qquad
\mu_2 = 
\sum_{{\bm x},{\bm y}} {\rm Tr}\,Q_{\bm x} Q_{\bm y}\,.
\label{binderdef}
\end{equation}
We consider correlations of the fundamental variable ${\bm z}_x$.
To obtain a gauge-invariant quantity, we consider correlations with
$\lambda$ strings, i.e., averages like
\begin{equation}
 \hbox{Re } \left\langle  
  \bar{\bm z}_{\bm x}\cdot {\bm z}_{\bm y} \prod_{\ell \in \cal C} \lambda_\ell
   \right\rangle,
\end{equation}
where the product extends over the link variables that belong to a
lattice path $\cal C$ connecting points $\bm x$ and $\bm y$. To
define quantities that have the correct FSS, the path ${\cal C}$ must
be chosen appropriately, as discussed in Ref.~\cite{APV-08}. Here, to
simplify the calculations, we only consider correlations between
points that belong to lattice straight lines.  We define
\begin{eqnarray}
G_V(d,L) ={1\over V} 
   \sum_{{\bm x}} 
  \hbox{Re}\left\langle
  \bar{\bm z}_{\bm x}\cdot {\bm z}_{{\bm x}+d \hat{\mu}}  
   \prod_{n=0}^{d-1} \lambda_{{\bm x}+n \hat{\mu},\mu}
   \right\rangle,
   \label{Gd}
\end{eqnarray}
where all coordinates should be taken modulo $L$ because of the
periodic boundary conditions.  Note that in the definition of $G_V$ we
average over all lattice sites ${\bm x}$ exploiting the translation
invariance of systems with periodic boundary conditions, and select a generic
lattice direction $\hat{\mu}$ (in our MC simulations we also average
over the three equivalent directions).  Note also that $G_V(0,L) = 1$
and that $G_V(L,L)$ is the average value $P(L)$ of the Polyakov loop,
\begin{equation}
   P(L) = {1\over V} \sum_{\bm x} 
   \hbox{Re } \left\langle \prod_{n=0}^{L-1} \lambda_{{\bm x}+n \hat{\mu},\mu}
   \right\rangle\,.
   \label{Polyakov}
\end{equation}
Finally, we consider the so-called Wilson loop
defined as
\begin{equation}
    W(m,L) = \hbox{Re } \left\langle \prod_{\ell\in \cal C}
    \lambda_\ell \right\rangle, 
    \label{Wilson}
\end{equation}
where the path $\cal C$ is a square of linear size $m$.

In the following we present a FSS analysis of the above observables,
for $N=2$ and $N=4$ and some values of $\beta_g>0$. In
Fig.~\ref{betac} we anticipate the resulting phase diagrams.  For both
$N=2$ and $4$, $\beta_c$ decreases as $\beta_g$ increases and,
eventually, it converges to the value appropriate for the $n$-vector
model with $n=4$ and 8, $\beta_c=0.233965(2)$~\cite{BFMM-96,CPRV-96}
and $\beta_c = 0.24084(1)$~\cite{DPV-15}.

As we shall discuss, our numerical data are consistent with a simple
scenario in which the nature of the transitions along the line
separating the confined and Higgs phases is unchanged for any finite
$\beta_g\ge 0$.  Therefore, for $N=2$ the phase transitions are continuous and
belong to the Heisenberg universality class as it occurs in the CP$^1$
model.  The O(4) critical behavior occurs only for $\beta_g$ strictly
equal to $\infty$. For $N=4$ instead, transitions are of first order,
except for $\beta_g=\infty$, where the system develops an O(8)
vector critical behavior.


\subsection{Continuous transitions for $N=2$}
\label{N2res}

As already mentioned, lattice versions of the three-dimensional CP$^1$
model undergo continuous transitions belonging to the Heisenberg
universality class, i.e., that of the standard $N=3$ vector model.
This has been also shown~\cite{PV-19} for model
(\ref{gllf}) with $\beta_g=0$ [$\beta_c = 0.7102(1)$ in this case].
On the other hand, for $\beta_g=\infty$ the model is equivalent 
to the standard O(4) vector model that has a continuous transition for 
\cite{BFMM-96,CPRV-96} $\beta_c=0.233965(2)$. 
As already inferred
from the RG flow of the AH continuum theory, the $\beta_g=\infty$ O(4)
critical behavior is expected to be unstable against perturbations
associated with nonzero values of $\beta_g^{-1}$. Therefore, the most
natural hypothesis is that the all transitions for finite $\beta_g\ge
0$ belong to the Heisenberg universality class.  However, a
substantial crossover from the O(4) to the O(3) behavior is expected
to characterize the transition for relatively large values of
$\beta_g$, $\beta_g\gtrsim 1$ say.

\begin{figure}[tbp]
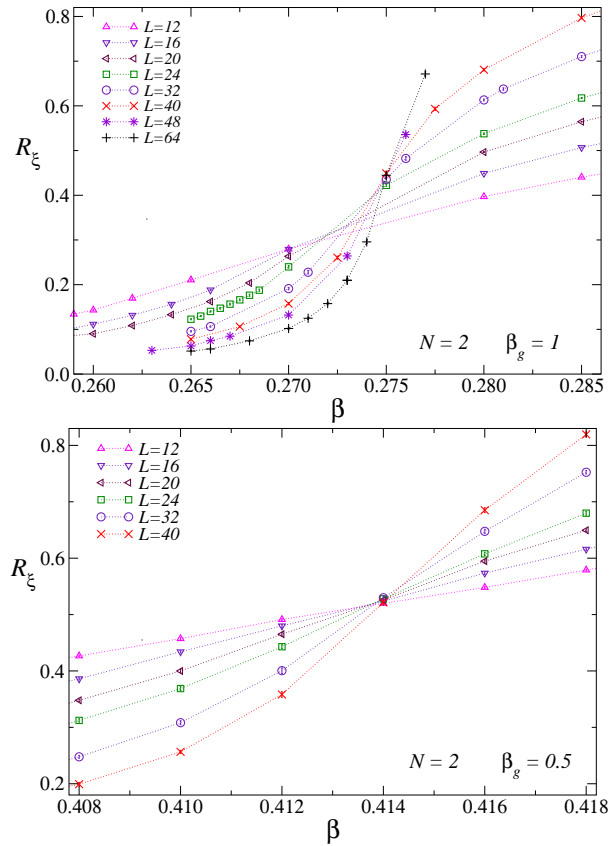

\includegraphics*[scale=\graphicscale]{rxi-n2-c1.eps}
\includegraphics*[scale=\graphicscale]{rxi-n2-c1o2.eps}
\caption{ $R_\xi$ versus $\beta$ for the $N=2$ AH lattice model, for
  $\beta_g=0.5$ (bottom) and $\beta_g=1$ (top).  In both cases the
  data for different values of $L$ show a crossing point, whose
  position provides an estimate of the critical point:
  $\beta_c=0.4145(5)$ and $\beta_c=0.276(1)$ for $\beta_g=0.5$ and
  $\beta_g=1$, respectively.  }
\label{rxi-n2}
\end{figure}

To provide evidence of this scenario, we perform MC simulations for
$\beta_g=0.5$ and 1. As in our previous work \cite{PV-19}, we study
the FSS behavior of the Binder parameter $U$ and of $R_\xi = \xi/L$.
At continuous transitions the FSS limit is obtained by taking
$\beta\to \beta_c$ and $L\to\infty$ keeping 
\begin{equation}
X \equiv (\beta-\beta_c)L^{1/\nu}
\label{Xdef}
\end{equation}
fixed. Any RG invariant quantity $R$, such
as $R_\xi\equiv \xi/L$ and $U$, is expected to asymptotically behave
as
\begin{eqnarray}
R(\beta,L) = f_R(X) + O(L^{-\omega})\,,
\label{rsca}
\end{eqnarray}
where $\omega>0$ is the leading scaling correction
exponent~\cite{PV-02}, and $f_R(X)$ is universal apart from a
normalization of its argument.  The function $f_R(X)$ only depends on
the shape of the lattice and on the boundary conditions.  In the case
of the Heisenberg universality we
have~\cite{GZ-98,PV-02,CHPRV-02,HV-11} $\nu=0.7117(5)$ and
$\omega=0.78(1)$.  As $R_\xi$ is monotonically increasing as a
function of $X$, Eq.~(\ref{rsca}) implies that
\begin{equation}
U = F(R_\xi) + O(L^{-\omega})\,,
\label{r12sca}
\end{equation}
where $F(x)$ is a universal scaling function.  As in our previous work
\cite{PV-19}, we will use Eq.~(\ref{r12sca}) to perform a direct check
of universality, because no model-dependent normalizations enter: If
two models belong to the same universality class, the data for both of
them should collapse onto the same curve as $L$ increases.  The only
difficulty in the approach is that one should be careful in
identifying corresponding operators in the two models.

To identify the correct operators, one may reason as follows. In the AH
lattice model the basic quantity that we consider is the local
operator (\ref{qdef}).  To identify the corresponding operator in the
Heisenberg model, we use the explicit relation between the CP$^1$ and
the O(3) vector model. Under the mapping, the parameter $U$ and $\xi$
correspond to the usual O(3) vector Binder parameter and correlation
length (i.e., computed from correlations of the fundamental spin
variable ${\bm s}_{\bm x}$). The mapping of the large-$\beta_g$ limit
of the AH lattice model into the O(4) vector model is instead more
complex and is discussed in detail in App.~\ref{FSSON} and in the
supplementary material \cite{suppl-mat}. The correspondence is not
trivial and $U$ is identified with a combination of suitably defined
O(4) tensor Binder parameters.

\begin{figure}[H]
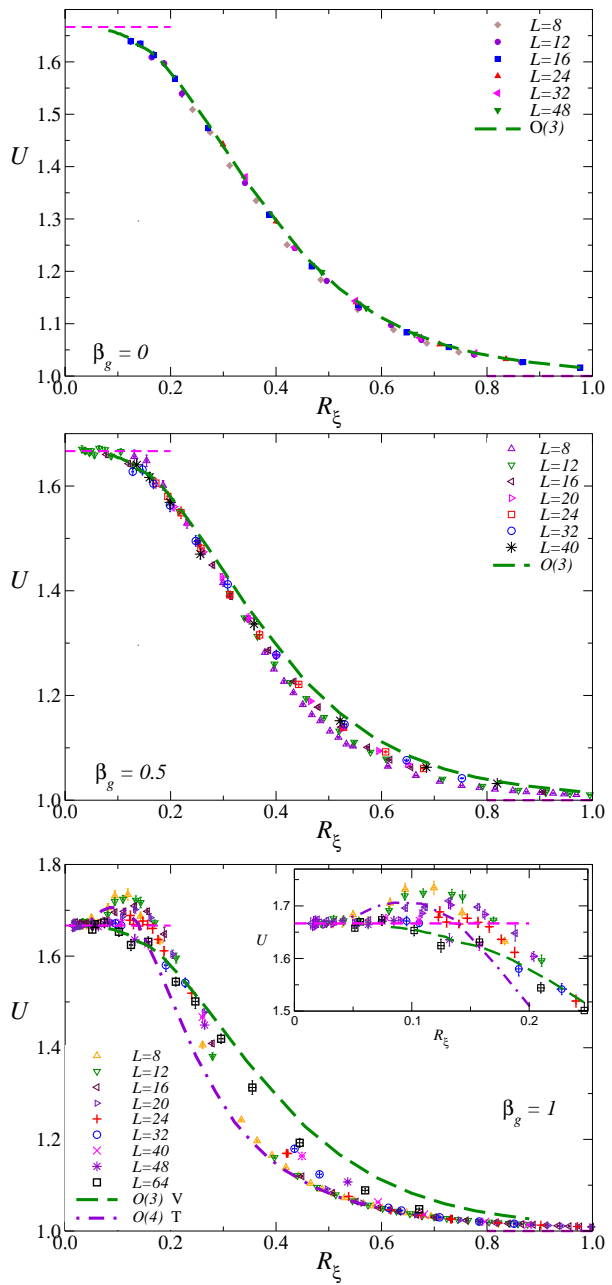

\includegraphics*[scale=\graphicscale]{bi-rxi-n2-c0.eps}
\includegraphics*[scale=\graphicscale]{bi-rxi-n2-c1o2.eps}
\includegraphics*[scale=\graphicscale]{bi-rxi-n2-c1.eps}
\caption{ The Binder parameter $U$ versus $R_\xi$ for the $N=2$ AH
  lattice model, and for $\beta_g=0$ (top, data from
  Ref.~\cite{PV-19}), $\beta_g=0.5$ (middle) and $\beta_g=1$ (bottom).
  In all panels the dashed line is the Heisenberg curve, as obtained
  from MC simulations of the O(3) vector model.  The dot-dashed line
  in the lower panel is the limiting curve for
  $\beta_g\to\infty$ (see App.~\ref{FSSON}).  The inset enlarges the
  region $R_\xi<0.25$, showing that the nonmotonic behavior that
  characterizes the small-size data (similar to the one of the O(4)
  curve) for $R_\xi\approx 0.1$ disappears with increasing $L$.  The
  horizontal dashed line shows the asymptotic value $U(R_\xi\to 0) =
  5/3$.  }
  \label{bi-rxi-n2}
\end{figure}

In Fig.~\ref{rxi-n2} we plot $R_\xi$ versus
$\beta$, for several values of $L$.  The data for different values of
the size $L$ show crossing points, which provide estimates of the
critical point: $\beta_c = 0.4145(5)$ and $\beta_c=0.276(1)$ for
$\beta_g=0.5$ and $\beta_g=1$, respectively.  Data are consistent with
a continuous transition. 

We now argue that the transitions are consistent with the expected
asymptotic Heisenberg behavior.  The best evidence is provided by the
plots of $U$ versus $R_\xi$, see Fig.~\ref{bi-rxi-n2}. In all panels
we report the data and the corresponding O(3) curve. If our simple
scenario is correct, the data for all values of $\beta_g$ must
approach the O(3) curve with increasing $L$.  For $\beta_g = 0$ we
observe very good agreement, as already discussed in
Ref.~\cite{PV-19}. For $\beta_g = 0.5$ convergence is slower,
indicating that scaling corrections increase with increasing
$\beta_g$.  For $R_\xi \lesssim 0.25$ we observe a good collapse of
the data, while in the opposite case, we observe a clear upward trend,
consistent with an asymptotic O(3) behavior.  For $\beta_g = 1$, for
small values of $L$ we observe significant differences between data
and O(3) curve.  These discrepancies can be explained by scaling
corrections. For $R_\xi\lesssim 0.25$, the results for $L=64$ fall on
top of the O(3) scaling curve, as predicted. For larger values of
$R_\xi$, crossover effects are stronger, but the trend of the data is
the expected one.

\begin{figure}[!ht]
\includegraphics*[scale=\graphicscale]{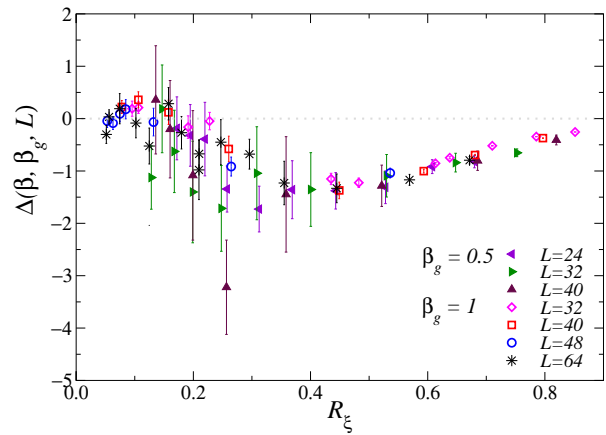}
\caption{The quantity $\Delta(\beta,\beta_g,L)$ defined in 
Eq.~(\ref{deltadef}) versus $R_\xi$. We report data for $\beta_g=1$ 
and 0.5 and several values of $L$.}
\label{differenze}
\end{figure}

To be more quantitative, let us note that, for large values of $L$,
the Binder parameter $U$ should behave as~\cite{PV-02}
\begin{equation}
U(\beta,\beta_g,L) = F(R_\xi) + a(\beta_g)\ G(R_\xi)\ L^{-\omega} + \ldots
\label{corrections}
\end{equation}
where $F(R_\xi)$ is the O(3) scaling function, $G(R_\xi)$ is a
universal function, and $a(\beta_g)$ is a constant that encodes the
$\beta_g$-dependent size of the leading scaling corrections decaying
as $L^{-\omega}$.  We have verified that our data for $\beta_g = 0.5$
and 1 are consistent with Eq.~(\ref{corrections}), if we take $\omega
= 0.78$ (the leading correction-to-scaling exponent in Heisenberg
systems \cite{GZ-98,PV-02,CHPRV-02,HV-11}) and $a(1)/a(0.5) \approx
5$.  This can be checked from Fig.~\ref{differenze}, where we report
\begin{equation}
\Delta(\beta,\beta_g,L) = 
    {1\over a(\beta_g)} L^\omega [U(\beta,\beta_g,L) - F(R_\xi)]\,,
\label{deltadef}    
\end{equation}
where $F(R_\xi)$ has been determined in the O(3) vector model, $\omega
= 0.78$, $a(1) = 5$, and $a(0.5) = 1$. All data reported in the figure
are consistent with a single scaling curve that would be identified
with the function $G(R_\xi)$ in Eq.~(\ref{corrections}).  The
existence of similar crossover effects for $\beta_g = 0.5$ and 1 is
another demonstration of universality.

It is interesting to note that the behavior of the data for small $L$
at $\beta_g=1$ can be interpreted as due to the presence of O(4) fixed
point that controls the critical behavior for $\beta_g\to\infty$.  In
the lower panel of Fig.~\ref{bi-rxi-n2}, we also plot the O(4) scaling
curve.  The data for $\beta_g=1$ apparently follow the O(4) curve for
small lattice sizes, and then move toward the O(3) curve with
increasing $L$.  In particular, note the nonmotonic behavior of the
data for small lattice sizes and $R_\xi\approx 0.1$, similar to the
one that characterizes the O(4) curve.  Such a behavior disappears
with increasing $L$ (see the inset in the lower panel of
Fig.~\ref{bi-rxi-n2}).

On the basis of the above numerical results we argue that the
finite-temperature transition is continuous for any finite $\beta_g\ge
0$ and belongs to the Heisenberg universality class.  However, for
relatively large values of $\beta_g$, say $\beta_g\gtrsim 1$, notable
crossover effects emerge.  They are apparently related to the presence
of the O(4) fixed point, which is the relevant one $\beta_g\to\infty$.
For large values of $\beta_g$, such effects may hide the asymptotic
Heisenberg behavior. For intermediate sizes, data are expected to show
an effective O(4) critical behavior, converging to the Heisenberg
behavior only for very large lattices.

\subsection{First-order transitions for $N=4$}
\label{N4res}

\begin{figure}[tbp]
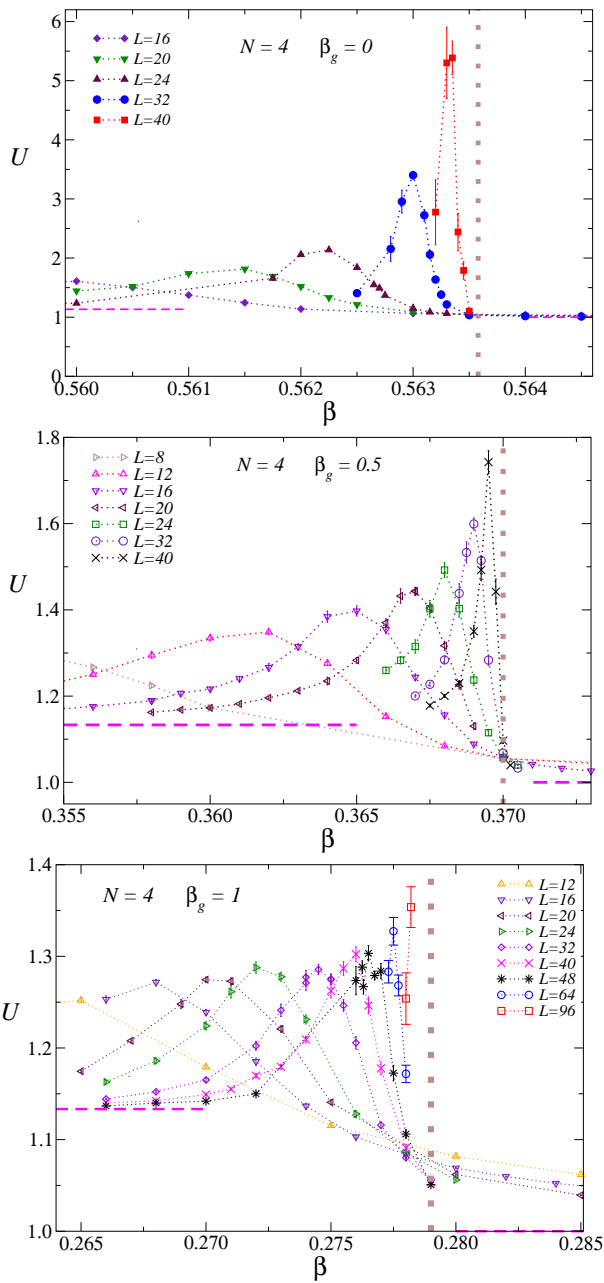

\includegraphics*[scale=\graphicscale]{u-n4-c0.eps}
\includegraphics*[scale=\graphicscale]{u-n4-c1o2.eps}
\includegraphics*[scale=\graphicscale]{u-n4-c1.eps}
\caption{Plot of the Binder parameter $U$ versus $\beta$, for
  $\beta_g=0$ (top, from Ref.~\cite{PV-19}), $\beta_g=1/2$ (middle),
  and $\beta_g=1$ (bottom), for $N=4$.  The vertical lines correspond
  to the estimates of the transition points.  The horizontal dashed
  lines show the values $U(\beta \to 0) = 17/15$ and
  $U(\beta\to \infty) = 1$.  } 
\label{U-N4}
\end{figure}

\begin{figure}[tbp]
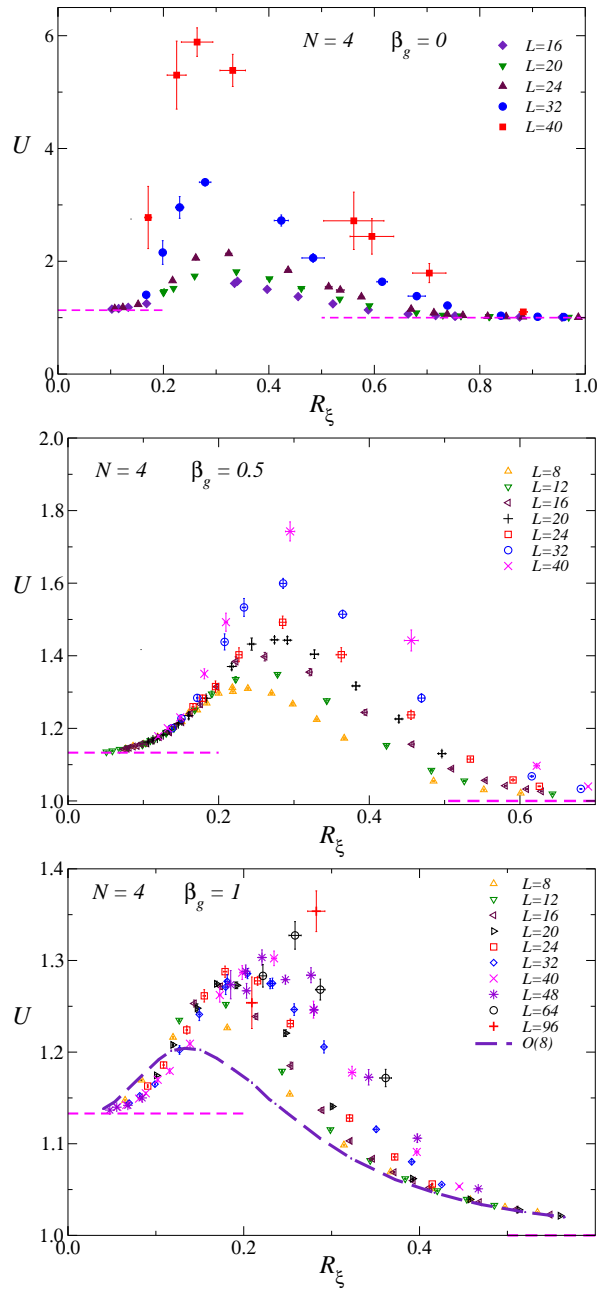

\includegraphics*[scale=\graphicscale]{bi-rxi-n4-c0.eps}
\includegraphics*[scale=\graphicscale]{bi-rxi-n4-c1o2.eps}
\includegraphics*[scale=\graphicscale]{bi-rxi-n4-c1.eps}
\caption{ The Binder parameter $U$ versus $R_\xi$ versus $\beta$, for
  $\beta_g=0$ (top, from Ref.~\cite{PV-19}), $\beta_g=1/2$ (middle),
  and $\beta_g=1$ (bottom) for $N=4$.  The dashed line in the lower
  panel is the O(8) limiting scaling curve, see App.~\ref{FSSON}.  The
  horizontal dashed lines show the asymptotic values $U(R_\xi\to 0) =
  17/15$ and $U(R_\xi\to \infty) = 1$.  }
\label{UvsR-N4}
\end{figure}

We now discuss the behavior of the $N=4$ AH lattice model, providing
evidence that the transitions along the line separating the Higgs and
confined phases are of first order for any finite $\beta_g$. Only when
$\beta_g$ is strictly infinity is the transition continuous: it
belongs to the O(8) vector universality class.

As shown in Ref.~\cite{PV-19}, the transition is of first order for
$\beta_g=0$.  To show that the nature of the transition is unchanged
for $\beta_g>0$, we first consider the specific heat and the Binder
parameter $U$. Both of them are expected to increase as the volume at
a first-order transition.  Indeed, according to the standard
phenomenological theory~\cite{CLB-86}, for a lattice of size $L$ there
exists a value $\beta_{{\rm max},C}(L)$ of $\beta$ where $C$ takes its
maximum value $C_{\rm max}(L)$, which asymptotically increases as
\begin{eqnarray}
&&C_{\rm max}(L) = V\left[ {1\over 4} \Delta_h^2 + O(1/V)\right]\,,
\label{cmaxsc}\\
&&\beta_{{\rm max},C}(L)-\beta_c\approx c\,V^{-1}\,, \label{betamax}
\end{eqnarray}
where $V=L^d$ and $\Delta_h$ is the latent heat
[defined as $\Delta_h = E(\beta\to\beta_c^+) - E(\beta\to\beta_c^-)$].
Analogously, the
behavior of the Binder parameter $U(\beta,L)$ is expected to show a
maximum $U_{\rm max}(L)$ at fixed $L$ (for sufficiently large $L$) at
$\beta = \beta_{{\rm max},U}(L) < \beta_c$
with~\cite{VRSB-93,CPPV-04,PV-19}
\begin{eqnarray}
&&U_{\rm max} \sim a\,V + O(1)\,,\label{umaxsca}\\ &&\beta_{{\rm
    max},U}(L) - \beta_c \approx b \,V^{-1}\,.
\label{bpeakU}
\end{eqnarray}
The previous relations are valid in the asymptotic limit and, for weak
transitions, require data on large lattices.  As we discussed in
Ref.~\cite{PV-19}, one can identify first-order transitions on
significantly smaller lattices from the analysis of the behavior of
the Binder parameter $U$. In the presence of a first-order transition,
one observes large violations of the scaling relation (\ref{r12sca})
for values of $L$ that are significantly smaller than those at which
relations (\ref{cmaxsc}) and (\ref{bpeakU}) hold. We will follow this
approach here, considering again two values of $\beta_g$, 0.5 and 1.

In Fig.~\ref{U-N4} we report numerical estimates of $U$ at $\beta_g=0$
(taken from Ref.~\cite{PV-19}), $\beta_g=0.5$, and
$\beta_g=1$. Clearly, the maximum $U_{\rm max}$ increases with
increasing $L$, as expected for a first-order transition. However,
with increasing $\beta_g$, the rate of increase becomes smaller,
indicating that the transition becomes weaker. The specific heat
behaves analogously.

To obtain a better evidence that the finite-size behavior is not
compatible with a continuous transition, we plot $U$ versus $R_\xi$,
see Fig.~\ref{UvsR-N4}. Data do not show any scaling behavior, as
expected at a first-order transition.

\begin{figure}[tbp]
\includegraphics*[scale=\graphicscale,angle=0]{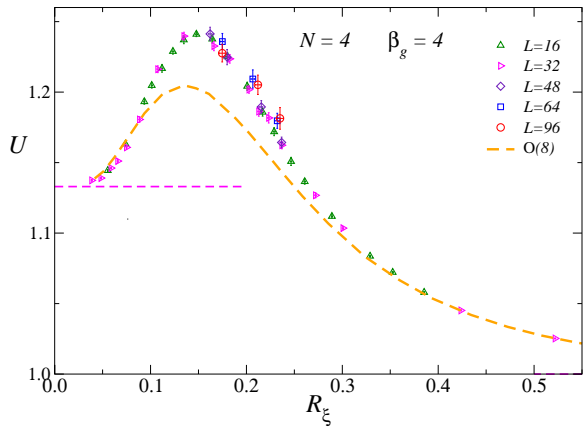}
\caption{ The Binder parameter $U$ versus $R_\xi$ for $\beta_g = 4$
  and $N=4$.  The full line connects the data with $L=16$.  The dashed
  line is the O(8) limiting scaling curve, see App.~\ref{FSSON}.  }
\label{U-N4-c4}
\end{figure}

Note that for $\beta_g = 1$ the small size data show an apparent
scaling behavior for small values of $R_\xi$ and small $L$, which may
lead to erroneous conclusions when limiting the FSS analysis to small
lattices (as in Ref.~\cite{KHI-11}).  To clarify the origin of the
transient effects, and understand whether they can be interpreted as
due to the O(8) fixed point that controls the behavior for $\beta_g =
\infty$, we have performed MC simulations for $\beta_g =
4$. This value is so large that, for our range of values of $L$, we do
not expect to observe effects related to the first-order nature of the
transition and therefore all data should be in the crossover
region. The analysis of $U$ as a function of $\beta$ allows us to
estimate $\beta_c = 0.2484(2)$, which is close to the O(8) value,
$\beta_c \approx 0.2408$. At the transition, gauge fields are
significantly ordered and indeed, the average value of the product of
the gauge fields along an elementary plaquette (a Wilson loop of size
1) is 0.95780(5) (for comparison, such a product is equal to 0.8235(5)
at the transition for $\beta_g = 1$). In Fig.~\ref{U-N4-c4} we report
$U$ versus $R_\xi$ and compare it with the O(8) curve. The numerical
data with $16\le L \le 48$ apparently fall onto a single scaling
curve, while the data corresponding to $L=64$ and 96 begin to show the
drift that characterizes the results at $\beta_g \le 1$ and which is
related to the asymptotic first-order nature of the transition.  The
apparent scaling curve for small values of $L$ is different from the
O(8) one, indicating that for $\beta_g = 4$ we are observing a sizable
contribution due the relevant operator that destabilizes the O(8)
fixed point for finite $\beta_g$. We can also infer from the
substantial stability of the results for $L\le 48$ that it has a very
small (positive) scaling dimensions $y$.  Indeed, close to the O(8)
fixed point, we expect
\begin{equation}
  U(\beta,\beta_g) = F(R_\xi,b(\beta_g,\beta) L^y) \,,
\end{equation}
where $b(\beta_g,\beta)$ is a nonuniversal amplitude, which vanishes
for $\beta_g \to \infty$. For each $\beta_g$ the crossover region is
the one in which $b(\beta_g,\beta_c) L^y \ll 1$. If this condition
holds, $U$ can be written as
\begin{equation}
  U(\beta,\beta_g) = F(R_\xi,0) + b(\beta_g,\beta) L^y G(R_\xi)\,,
\end{equation}
where the first term is the O(8) scaling function.  This equation
would imply that the deviations from the O(8) behavior scale, at least
for $\beta_g$ very large, as $L^y$. Our results therefore imply that
$y$ should be small enough, so that $L^y$ does not change
significantly as $L$ varies from 16 to 48.

In conclusion, the numerical results favor a phase diagram based on a
first-order transition line for $\beta_g>0$, starting from the
first-order transition of the CP$^3$ models, corresponding to
$\beta_g=0$.  With increasing $\beta_g$ the first-order transition
becomes weaker.  We observe substantial crossover phenomena for
$\beta_g\gtrsim 1$.  They may be explained in terms of the O(8) fixed
point controlling the behavior for $\beta_g\to\infty$, perturbed by a
relevant operator with a relatively small scaling dimension.

\subsection{Vector and gauge observables} \label{Vector}

In the previous sections we discussed the behavior of quantities
defined in terms of the gauge-invariant order parameter $Q_{\bm
  x}^{ab}$. Here we discuss instead the vector correlation function
(\ref{Gd}) and the gauge observables (\ref{Polyakov}) and
(\ref{Wilson}). We focus on $N=4$.

\begin{figure}[tbp]
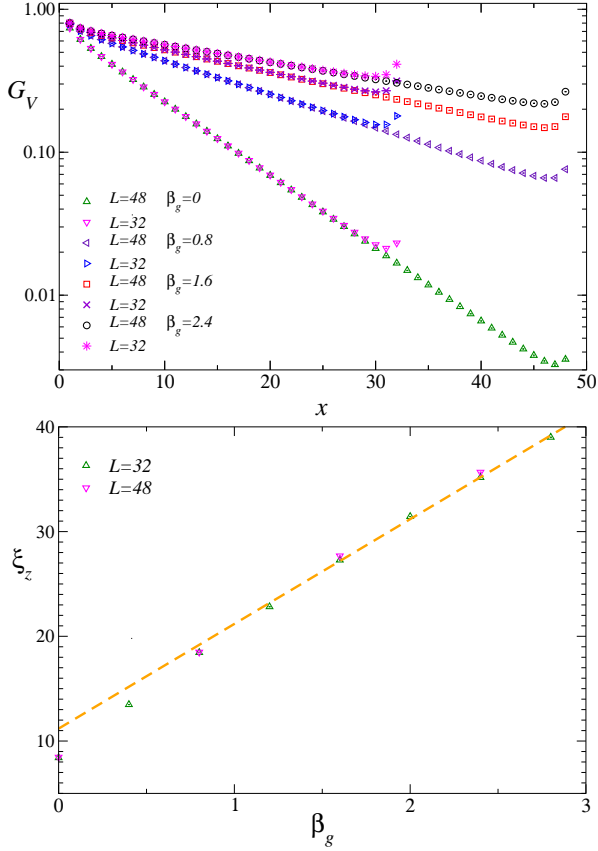

\includegraphics*[scale=\graphicscale,angle=0]{gdcrit-beta0p8.eps}
\includegraphics*[scale=\graphicscale,angle=0]{xiz-beta0p8.eps}
\caption{Top: Vector correlation function $G_V(x,L)$ versus $x$ for
  $\beta=0.8$ and several values fo $\beta_g$. Bottom: Vector
  correlation length $\xi_z$ as a function of $\beta_g$ for $\beta =
  0.8$. The line shows that $\xi_z$ scales as $\beta_g$ for large
  $\beta_g$ (the parameters have been determined by performing a
  linear fit of the data with $\beta_g \ge 1.6$).  For $x=L$, $G_V(L,L)$ 
  corresponds to the average of the Polyakov loop.
}
\label{Gdbeta0p8}
\end{figure}

\begin{figure}[tbp]
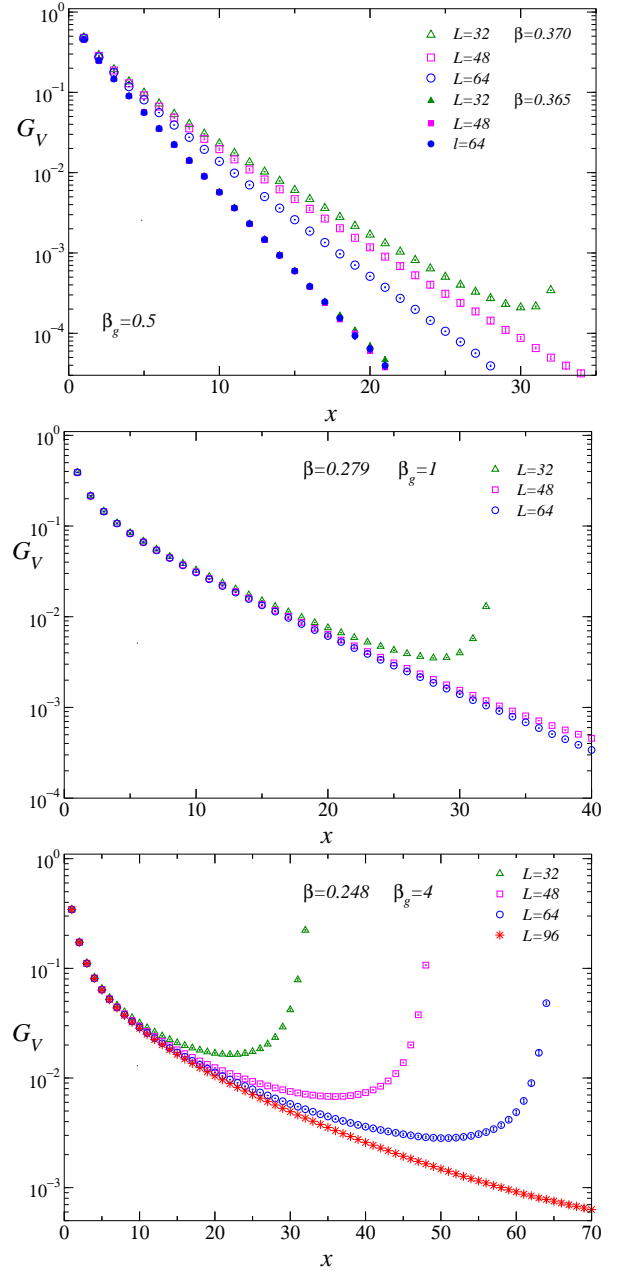

\includegraphics*[scale=\graphicscale,angle=0]{gdcrit-c0p5.eps}
\includegraphics*[scale=\graphicscale,angle=0]{gdcrit-c1.eps}
\includegraphics*[scale=\graphicscale,angle=0]{gdcrit-c4.eps}
\caption{The vector correlation function $G_V(x,L)$ in the critical
  region for $\beta_g = 0.5$ (top), 1 (middle) and 4 (bottom). For
  $\beta_g = 0.5$ we report the estimate for $\beta = 0.365$
  (high-temperature phase) and for $\beta = 0.370$ (low-temperature
  phase). For $\beta_g = 1$ and 4, we report an effective estimate of
  the correlation function in the coexistence region (see text for a
  discussion), computed at $\beta = 0.279$ for $\beta_g = 1$, and at
  $\beta = 0.248$ for $\beta_g = 4$.  }
\label{Gdcrit}
\end{figure}

Let us first discuss their behavior in the two different phases. In
the high-temperature phase $\beta < \beta_c$, we find that the
correlation function can be approximated as ($x > 0$)
\begin{equation}
  G_V(x,L) = A e^{-x/\xi_z}\,,
\label{Gd-exp}
\end{equation}
as soon as $x$ is 2 or 3. Moreover, for small $\beta$, $\xi_z$ is very
little dependent on $\beta_g$.  For instance, for $\beta = 0.1$, the
strong-coupling behavior $G_V(x,L) \sim (N\beta)^x$ holds quite precisely
for all values of $\beta_g$. Wilson loops behave in a very similar
fashion. We find $W(m,L) \approx B \exp(-4 m/\xi_w)$ with $\xi_w
\approx \xi_z$ as long as $m\gtrsim 2$.  Clearly, in the
high-temperature phase a single gauge mode controls the behavior of
all observables that involve gauge degrees of freedom.

The behavior in the low-temperature phase is analogous.  The
correlation function $G_V(x,L)$ behaves as in Eq.~(\ref{Gd-exp}); see
the upper panel of Fig.~\ref{Gdbeta0p8} for results at $\beta = 0.8$.
Moreover, Polyakov and Wilson loops satisfy
\begin{equation}
P(L) = A e^{-L/\xi_z}\,, \qquad W(m) = B e^{-4 m /\xi_z}\,,
\label{P-exp}
\end{equation}
with the same correlation length and $A,B\approx 1$.
Fig.~\ref{Gdbeta0p8} also shows that $G_V(x,L)$ has a very precise
exponential decay even when $\xi_z \gtrsim L$. Clearly, it couples to
a single isolated mode and hence there are no corrections to the
leading exponential behavior.  In this phase the correlation length
increases with $\beta_g$ (see the lower panel of
Fig.~\ref{Gdbeta0p8}): in agreement with perturbation theory, it
scales linearly with $\beta_g$ in the limit $\beta_g \to \infty$. Note
the $\xi_z$ is also expected to diverge in the limit $\beta\to \infty$
at fixed $\beta_g$. Indeed, for $\beta\to \infty$, the relevant
configurations are those that minimize the Hamiltonian term that
depends on the fields ${\bm z}$. If we perform a local minimization on
each link, we find the constraint
\begin{equation}
    z_{{\bm x} + \hat{\mu}} = \bar\lambda_{{\bm x},\mu} z_{\bm x}\,.
\end{equation}
This constraint can be satisfied simultaneously on the four links
belonging to a plaquette only if the product of the gauge fields along
the plaquette is 1. Analogously, the constraint is satisfied on the
links that belong to a loop that wraps around the lattice only if the
Polyakov operator is 1. It follows that gauge configurations are
trivial---$\lambda_{{\bm x},\mu}$ is 1 on all links modulo gauge
transformations---and $\xi_z$ is infinite in this limit.

These results for the gauge observables indicate that gauge and vector
observables are noncritical in both phases and that their behavior is
analogous for small and large values of $\beta$.  Only the limit
$\beta_g\to \infty$ distinguishes the two sectors. If
$\beta_{c,\infty}$ is the transition point for $\beta_g\to \infty$
[therefore in the O($2N$) theory], for $\beta_g\to \infty$, the
correlation length $\xi_z$ is finite for $\beta < \beta_{c,\infty}$
and infinite in the opposite case. This guarantees that vector
correlations are critical in the O($2N$) theory with a finite
low-temperature magnetization.  But this only occurs for $\beta_g$
strictly equal to infinity. For finite $\beta_g$, only $Q$
correlations display criticality.

Finally, let us discuss the behavior of vector and gauge quantities
along the transition line.  For $N=4$, as we are dealing with
first-order transitions, we expect $G_V(x,L)$ to depend on the phase
one considers. In the CP$^3$ model ($\beta_g = 0$) the transition is
strong and therefore the high-temperature (HT) and low-temperature
(LT) correlation functions can be easily computed by fixing $\beta$ in
the coexistence region and starting the simulation from a random or an
ordered configuration.  We find that in both cases the correlation
function decays very rapidly and estimate $\xi_z \approx 1.9$ and
$\xi_z \approx 1.7$ in the LT and HT phase, respectively.  Clearly,
vector modes are not critical. Similar results hold for the CP$^1$ and
CP$^2$ models. In the first case, we obtain $\xi_z \approx 2.2 $.  For
$N=3$ the transition is so weak that we cannot identify the two phases
and we are only able to compute an effective correlation function,
which is a linear combination of those appropriate for the two phases.
This quantity still allows us to compute the largest of the two
correlation lengths, i.e., $\xi_z$ in the LT phase, obtaining $\xi_z
\approx 2.2$.  Results for finite $\beta_g$ are reported in
Fig.~\ref{Gdcrit}. The first distinctive feature is that, for the
cases we consider, the correlation function does not behave as a
single exponential, although an exponential behavior sets when $x \gg
\xi_z$.  Clearly, at the critical point several modes are playing an
important role and an exponential behavior is only observed when
$\xi_z$ is significantly less than $L$.  Second, the correlation
length $\xi_z$ increases with increasing $\beta_g$ along the
transition line. For $\beta_g = 0.5$ we obtain $\xi_z \approx 3.8$,
3.6, 3.1 for $L=32$, 48, and 64 in the LT phase (runs at $\beta =
0.370$ with ordered start).  In the HT phase (runs at $\beta = 0.365$)
we obtain $\xi_z \approx 2.2$ with a small $L$ dependence.  Although
$\xi_z$ is small, it is larger than the value it takes in the CP$^3$
model, i.e. the AH model with $\beta_g=0$.  For $\beta_g = 1$, we are
not able to distinguish the two phases and, therefore, we only compute
the LT estimate of $\xi_z$. Results for $L=32,48,64$ essentially agree
and give $\xi_z \approx 6.9$. This estimate is confirmed by the
analysis of the Polyakov loop.  A fit to Eq.~(\ref{P-exp}) gives
$\xi_z = 6.9(1)$, in very good agreement with the results obtained
from $G_V(x)$.  For $\beta_g = 4$, even for $L = 96$ we are not yet in
the regime in which one can reliably identify a range of distances in
which the correlation function decays exponentially. If we fit the
correlation function to Eq.~(\ref{Gd-exp}) in the range $L/3 \le 2
L/3$, we obtain $\xi_z = 17.1(1)$ and 17.6(1) for $L=64$ and 96,
respectively.  The analysis of the Polyakov loop gives a somewhat
larger value $\xi_z = 20.9(3)$. Whatever the exact asymptotic result is,
data confirm that, for $\beta_g = 4$, we are deep in the crossover
region, where vector and gauge excitations compete with
gauge-invariant excitations associated with $Q_{\bm x}$ (for
comparison note that $\xi = 20.3(3)$ for $L=96$).  These results
provide us a physical explanation of the crossover effects we
observe. The asymptotic first-order behavior is only observed when the
correlation length $\xi(L)$ at the transition point is significantly
larger than $\xi_z$. When $\xi(L) \sim \xi_z$ we observe an apparent
scaling behavior in which both the (gauge-field independent) degrees
of freedom associated with $Q$ and the (gauge-field dependent) ones,
that are encoded in the gauge observables and in the vector
correlations, are both relevant.

\section{Conclusions} \label{Conclusions}

We have studied the phase diagram and critical behavior of
muticomponent AH lattice models, in which an $N$-component complex
field ${\bm z}_{\bm x}$ is coupled to quantum electrodynamics.  We
consider the compact Wilson formulation of Abelian lattice gauge
theories in which the fundamental gauge fields are complex numbers of
unit modulus, see Eq.~(\ref{gllf}).  For the scalar fields, we
consider the unit-length limit and fix $|{\bm z}_{\bm x}|^2 =
1$. Finally, we fix $q=1$ for the charge of the matter fields. We
focus on systems with a small number of components, considering $N=2$
and $N=4$.

We investigate the phase diagram of the model as a function of the
couplings $\beta$ and $\beta_g$. The phase diagram is characterized by
two phases: a low-temperature phase (large $\beta$) in which the order
parameter $Q^{ab}$ condenses, and a high-temperature disordered phase
(small $\beta$).  The gauge coupling does not play any particular role
in the two phases: gauge observables and vector observables do not
show long-range correlations for any finite $\beta$ and $\beta_g$. The
two phases are separated by a transition line that connects the
CP$^{N-1}$ transition point ($\beta_g = 0$) with the O($2N$)
transition point ($\beta_g = \infty$).  Concerning the nature of the
transition line, our numerical data are consistent with a simple
scenario, in which the nature of the transition line is independent of
$\beta_g$.  Therefore, we predict Heisenberg critical behaviors along
the whole line for $N=2$, and first-order transitions for $N=4$. Note
that, for $\beta_g\to \infty$ the model becomes equivalent to the
O($2N$) vector model, and therefore one expects strong crossover
effects controlled by the O($2N$) fixed point.  These crossover
effects are related to the presence of a second length scale $\xi_z$
associated with the vector correlations, which is finite for any
$\beta_g$ and diverges in the limit $\beta_g\to\infty$ in the whole
low-temperature phase. 

The scenario supported by our numerical data is fully consistent with
the LGW approach that assumes a gauge-invariant order parameter. On
the other hand, at least for $N=2$, it disagrees with the
$\varepsilon$-expansion predictions obtained using the standard
continuum AH model: for $N=2$ this approach does not predict a
continuous transition.  Numerical results allow us to understand why
the LGW approach is more appropriate than the continuum AH model for
these values of $N$.  At the transition (for both $N=2$ and $N=4$)
only correlations of the gauge-invariant operator $Q^{ab}$ display
long-range order. Gauge modes represent a background that gives only
rise to crossover effects and indeed, the asymptotic behavior sets in
only when the correlation length of the gauge fluctuations is
negligible compared to that of the $Q$ correlations.  It is important
to note that a LGW approach based on a gauge-invariant order parameter
has also been applied to the study of phase transitions in the
presence of nonabelian gauge symmetries, and, in particular, to the
study of the finite-temperature transition of hadronic matter as
described by the theory of strong interactions, quantum chromodynamics
\cite{PW-84,BPV-03,PV-13}. Our results for the AH lattice model lend
support to the correctness of the approach and of the predictions
obtained.

We expect that AH lattice models with higher (but not too large)
values of $N$ have a phase diagram similar to the one obtained for
$N=4$, with a first-order transition line separating the ordered and
disordered phases. The phase diagram may change for large values of
$N$. In this regime, the system may undergo continuous transitions
controlled by the stable fixed point of the continuum AH model.  This
issue requires additional investigations.

It is important to stress that we have considered here a compact
version of electrodynamics. Other models of interest in
condensed-matter physics consider complex fields (spinons) coupled to
noncompact electrodynamics \cite{SVBSF-04,SBSVF-05,SBSVF-04}. Such a
model may have a different critical behavior due the suppression of
monopoles \cite{MHSF-08,BMK-13,SP-15,DMPS-15}.  Numerical studies have
identified the transition, but at present there is no consensus on its
order.  The same is true for loop models which supposedly belong to
the same universality class (if it exists), see, e.g.,
Refs.~\cite{MV-04,KMPST-08,NCSOS-11,NCSOS-13,NCSOS-15}.  Clearly,
additional work is needed to settle the question.

\appendix

\section{Finite-size scaling behavior for $\beta_g\to \infty$}
\label{FSSON}

In this Appendix we discuss the limit $\beta_g \to \infty$ of the model.
In this limit, the gauge part of the Hamiltonian becomes trivial 
and we obtain 
\begin{equation} 
\lambda_{{\bm x},{\mu}} \,\lambda_{{\bm x}+\hat{\mu},{\nu}}
\,\bar{\lambda}_{{\bm x}+\hat{\nu},{\mu}}  \,
\bar{\lambda}_{{\bm x},{\nu}} = 1
\end{equation}
on every lattice plaquette. We consider a finite lattice with periodic
boundary conditions and further assume that the Polyakov loops order
in the same limit. If this occurs (we discuss this issue in
Sec.~\ref{Vector}), we can set $\lambda_{x,\mu} = 1$ on each lattice
link. Therefore, for $\beta_g\to\infty$, the Hamiltonian becomes
simply
\begin{equation} 
H = - \beta N
\sum_{{\bm x}, \mu}
\left( \bar{\bm{z}}_{\bm x} 
\cdot {\bm z}_{{\bm x}+\hat\mu} 
+ {\rm c.c.}\right)\,.
\end{equation}
We now define a $2N$-dimensional unit real vector ${\bm s}_{\bm x}$ by setting
\begin{equation}
  z^a_{\bm x} = s^a_{\bm x} + i s^{a+N}_{\bm x},\,
\end{equation}
$a=1,\ldots,N$. In terms of this new field the Hamiltonian becomes 
\begin{equation}
H_V = - n \beta 
\sum_{{\bm x}, \mu} \bm{s}_{\bm x} 
\cdot {\bm s}_{{\bm x}+\hat\mu} 
\label{Ovmo}
\end{equation}
with $n = 2 N$, which is the Hamiltonian of the $n$-vector model. We
have therefore an enlargement of the global symmetry: the model is now
invariant under O($2N$) transformations.

\begin{figure}[tbp]
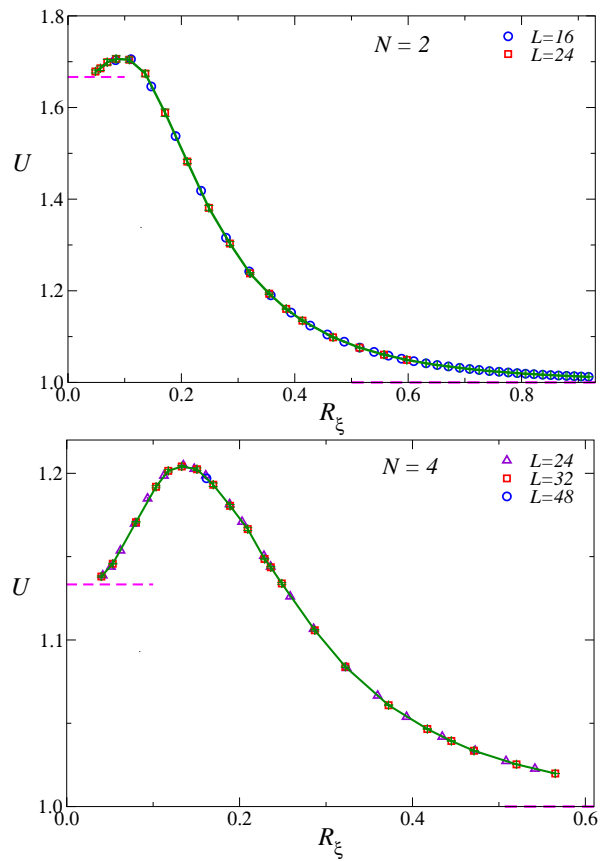

\includegraphics*[scale=\graphicscale]{bi-rxi-o4.eps}
\includegraphics*[scale=\graphicscale]{bi-rxi-o8.eps}
\caption{Scaling functions for the Binder parameter $U$ versus
  $R_{\xi}\equiv\xi/L$ for $N=2$ (top) and $N=4$ (bottom), in the
  large-$\beta_g$ limit, where the model is equivalent to the
  $n$-vector model with $n = 2N$. The are obtained by using
  Eq.~(\ref{U-ON}) and numerical results for the O(4) and O(8) models,
  respectively. Scaling corrections are tiny: the dashed lines
  represent a good approximation of the asymptotic FSS curve.  The
  horizontal dashed lines show the asymptotic values 
$U(R_\xi\to 0) =  5/3$ and $U(R_\xi\to \infty) = 1$ for $N=2$,    
$U(R_\xi\to 0) =  17/15$ and $U(R_\xi\to \infty) = 1$ for $N=4$.   
 }
\label{bi-rxi-oN}
\end{figure}

Since the model becomes O($2N$) invariant, it is useful to rewrite
CP$^{N-1}$ observables in terms of explicitly O($2N$) invariant
quantities that can be determined directly in the O($2N$) theory. The
basic CP$^{N-1}$ variable $Q_{\bm x}^{ab}$ can be rewritten in terms
of the tensor (spin-2) operator of O($2N$) theory:
\begin{equation}
T_{\bm x}^{\alpha\beta} = s_{\bm x}^\alpha s_{\bm x}^\beta - 
   {1\over 2N} \delta^{\alpha \beta}\,,
\end{equation}
where $\alpha,\beta = 1,\ldots 2N$. The relation is not trivial,
\begin{equation}
Q_{\bm x}^{ab} = T^{ab}_{\bm x} +  T^{a+N,b+N}_{\bm x} + 
   i T^{a,b+N}_{\bm x} - i T^{a+N,b}_{\bm x}\,,
\label{Q-vs-T}
\end{equation}
which implies that $Q$-correlations are not trivially related to $T$
correlations in the vector model.  Using relation (\ref{Q-vs-T}), we
can express the CP$^{N-1}$ correlation function $G({\bm x})$ in terms
of the tensor correlation function
\begin{equation}
G_T({\bm x}) = \sum_{\alpha\beta} 
   \langle T^{\alpha\beta}_{\bm 0} T^{\alpha\beta}_{\bm x} \rangle.
\end{equation}
Using the O($2N$) invariance of the model we obtain
\begin{equation}
G({\bm x}) = {2 (N-1)\over 2N-1} G_T({\bm x})\,,
\end{equation}
which implies that the CP$^{N-1}$ correlation length can be identified
with the tensor correlation length defined in the $2N$-vector model
using $G_T({\bm x})$ and Eq.~(\ref{xidefpb}). The Binder parameter $U$
can also be expressed in terms of analogous quantities defined in the
vector model. The relation is more complex and is discussed in detail
in the supplementary material \cite{suppl-mat}. In the $n$-vector
model we define
\begin{equation}
U_{4T,a} = 
{\langle \nu_2^2\rangle \over \langle \nu_2 \rangle^2} \,, \qquad
U_{4T,b} = 
{\langle \nu_4\rangle \over \langle \nu_2 \rangle^2} \,, 
\end{equation}
where 
\begin{eqnarray}
\nu_2 &=& 
\sum_{{\bm x}{\bm y}} {\rm Tr}\,T_{\bm x} T_{\bm y}\, , \\
\nu_4 &=& 
\sum_{{\bm x}{\bm y}{\bm z}{\bm t}} {\rm Tr}\,T_{\bm x} T_{\bm y}
   T_{\bm z} T_{\bm t}\, . 
\end{eqnarray}
A long calculation gives the relation
\begin{eqnarray}
U = {N (2 N - 1)  \left[(2 N^2 - 5 N + 4) U_{4T,a}  -    2 U_{4T,b} \right]
\over (N-1)^2 (2 N + 1) (2 N - 3)}, 
\end{eqnarray}
where $U_{4T,a}$ and $U_{4T,b}$ are computed in the $n$-vector model
with $n=2N$.  For $N=2$ and $N=4$, the two cases of relevance in this
work, we obtain:
\begin{eqnarray}
U &=& {12\over 5} (U_{4T,a} - U_{4T,b})\,, \nonumber \\
U &=& {448\over 405} U_{4T,a} - {56\over 405} U_{4T,b}\,,
\label{U-ON}
\end{eqnarray}
respectively.  To obtain the scaling functions associated with $U$ in
the large-$\beta_g$ limit for $N=2$ and 4, we have therefore performed
simulations in the $n$-vector model with $n = 4$ and 8, we have
computed the tensor Binder parameters $U_{4T,a}$ and $U_{4T,b}$, and
we have applied Eq.~(\ref{U-ON}).  Details on the numerical
simulations are reported in the supplementary material
\cite{suppl-mat}.  In Fig.~\ref{bi-rxi-oN} we show the resulting curve
for $N=2$ and $N=4$.

\clearpage

\begin{widetext}

\centerline{\bf \Large Supplementary material for }
\centerline{\bf \Large   ``Multicomponent compact Abelian-Higgs lattice models"}
\end{widetext}

\subsection{Scaling functions in the $N$ vector model}

We consider the three-dimensional $N$-vector model on a cubic lattice. We define an
$N$-dimensional real spin vector $s^\alpha_{\bm x}$ on each lattice site 
and consider the Hamiltonian
\begin{equation}
{\cal H} = - \sum_{\langle {\bm xy}\rangle} 
    {\bm s}_{\bm x}\cdot {\bm s}_{\bm y}.
\label{H-suppl}
\end{equation}
The sum in Eq.~(\ref{H-suppl}) extends over all lattice nearest-neighbor pairs
$\langle {\bm xy}\rangle$. 
In this supplementary material we investigate the critical 
behavior of tensor observables, defined in terms of the tensor (matrix) field
\begin{equation}
T^{\alpha\beta}_{\bm x} =
{s}^\alpha_{\bm x} s^\beta_{\bm x} - {1\over N} \delta^{\alpha\beta}.
\end{equation}
We define the tensor correlation function 
\begin{equation}
G_T({\bm x}) = 
\sum_{\alpha\beta} \langle T^{\alpha\beta}_{\bm 0} T^{\alpha\beta}_{\bm x} 
\rangle = 
\langle \hbox{Tr}\,(T_{\bm 0} T_{\bm x}) \rangle,
\label{GT-suppl}
\end{equation}
where ``Tr" is the trace over the $O(N)$ indices, and the corresponding correlation 
length 
\begin{equation}
\xi_T^2 = {1\over 4 \sin^2 (\pi/L)} 
   {\widetilde{G}_T({\bm 0}) - \widetilde{G}_T({\bm p}_m) \over 
  \widetilde{G}_T({\bm p}_m)}.
\label{xiT-suppl}
\end{equation}
Here 
$\widetilde{G}_T({\bm p})= \sum_{\bm x} G_T({\bm x}) e^{i{\bm p}\cdot {\bm x}}$ 
and ${\bm p}_m = (2\pi/L,0,0)$. 
Definitions (\ref{GT-suppl}) and (\ref{xiT-suppl}) 
are the obvious generalizations 
of those used for spin-spin correlations. In this case 
one considers the vector correlation function 
\begin{equation}
G_V({\bm x}) = \langle {\bm s}_{\bm 0} \cdot {\bm s}_{\bm x} \rangle ,
\end{equation}
and the vector correlation length $\xi_V$ defined using Eq.~(\ref{xiT-suppl}) 
and $G_V({\bm x})$.

Beside the correlation length, we also consider renormalization-group invariant ratios 
(collectively named Binder parameters) defined in terms of 
\begin{equation}
\Theta^{\alpha\beta}  = \sum_{\bm x} T_{\bm x}^{\alpha \beta}.
\label{defTheta-suppl}
\end{equation}
We define 
\begin{equation}
U_{3T} = {\langle \hbox{Tr } \Theta^3 \rangle \over \langle \hbox{Tr } 
   \Theta^2\rangle^{3/2}},
\end{equation}
\begin{equation}
U_{4T,a} = {\langle [\hbox{Tr } \Theta^2]^2\rangle  \over 
      \langle \hbox{Tr }\Theta^2\rangle^2 } \qquad
U_{4T,b} = {\langle \hbox{Tr } \Theta^4 \rangle \over \langle \hbox{Tr }
\Theta^2 \rangle^2}\; .
\end{equation}
These quantities are not independent for $N=2$ and 3. 
The relation $U_{4T,b} = U_{4T,a}/2$ holds for $N=2,3$, while $U_3 = 0$
for $N=2$.

We can easily predict the value the Binder parameters take in the high-temperature 
phase. The cubic parameter $U_{3T}$ vanishes, while
\begin{equation}
U_{4T,a} = {N^2 + N + 2 \over (N+2) (N - 1)} \quad
U_{4T,b} = {2 N^2 + 3 N - 6 \over N (N+2) (N - 1)}.
\end{equation}
In the low-temperature phase we obtain instead
\begin{equation}
U_{3T} = {N-2\over \sqrt{N(N-1)} },
\end{equation}
\begin{equation}
U_{4T,a} = 1 \qquad U_{4T,b} = {N^2 - 3 N + 3\over N (N-1)}.
\end{equation}
Numerical results for $N=3,4,5,8$ 
are reported in Table~\ref{Rstar-HT-LT-suppl}.

\begin{table}
\caption{Estimates of the Binder parameters in the high-temperature (HT)
and low-temperature (LT) phase. $U_{3T} = 0$ in the HT phase.}
\label{Rstar-HT-LT-suppl}
\begin{tabular}{lccccc}
\hline\hline
$N$  & \multicolumn{2}{c}{HT} & \multicolumn{3}{c}{LT} \\
\hline
   & $U_{4T,a}$ & $U_{4T,b}$ & $U_{3T}$ & $U_{4T,a}$ & $U_{4T,b}$ \\
\hline
3  & 1.400  & 0.700 & 0.408 & 1 & 0.500 \\
4  & 1.222  & 0.528 & 0.577 & 1 & 0.583 \\
5  & 1.143  & 0.421 & 0.671 & 1 & 0.650 \\
8  & 1.057  & 0.261 & 0.802 & 1 & 0.768 \\
$\infty$ & 1& 0 & 1 & 1 & 1 \\
\hline\hline
\end{tabular}
\end{table}

\begin{figure}[tbp]
\includegraphics*[scale=\graphicscale,angle=-90]{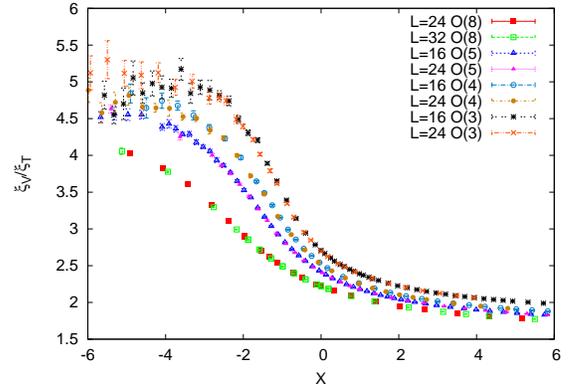} 
\caption{
Plot of $\xi_V/\xi_T$ as a function of $X = (\beta - \beta_c) L^{1/\nu}$ 
for $N=3,4,5$ and 8.}
\label{RatioxivsX-suppl}
\end{figure}

Close to the critical point, for $L\to \infty$, 
renormalization-group invariant quantities $R$ satisfy the scaling law
\begin{equation}
R(\beta,L) = f_R(X) + O(L^{-\omega}) \qquad 
X = (\beta - \beta_c) L^{1/\nu}.
\label{R-vsX-suppl}
\end{equation}
Here $\beta_c$ is the critical-point position, $\nu$ the correlation-length exponent
and $\omega$ is the exponent that controls scaling corrections (in $N$ vector models 
it varies between 0.8 and 1, see Ref.~\cite{PV-02-suppl}). 
The function $f_R(X)$ is universal
apart from a rescaling of its argument. 
In particular, its value $R^* = f_R(0)$ at the critical
point is universal. As we have done in the paper, 
we can also parametrize the scaling behavior
by using a specific quantity $R$. Here we use the 
ratio $R_{\xi,T} = \xi_T/L$, rewriting 
Eq.~(\ref{R-vsX-suppl}) as 
\begin{equation}
R(\beta,L) = g_R(\xi_T/L) + O(L^{-\omega}).
\label{R-vsxiTsuL-suppl}
\end{equation}
The function $g_R(x)$ is universal.

\begin{figure*}[tbp]
\begin{tabular}{cc} 
\includegraphics*[scale=\graphicscale,angle=-90]{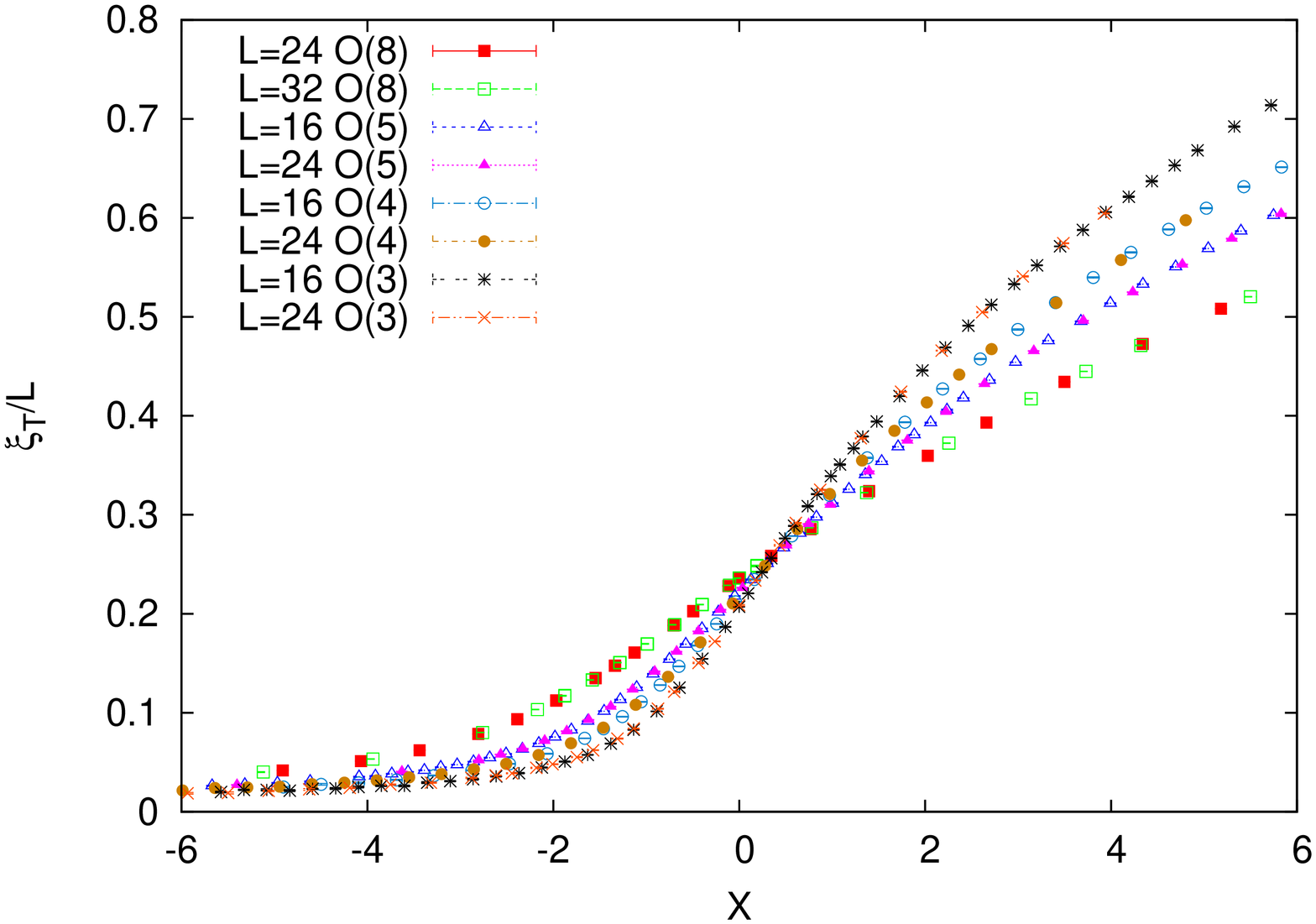} &
\includegraphics*[scale=\graphicscale,angle=-90]{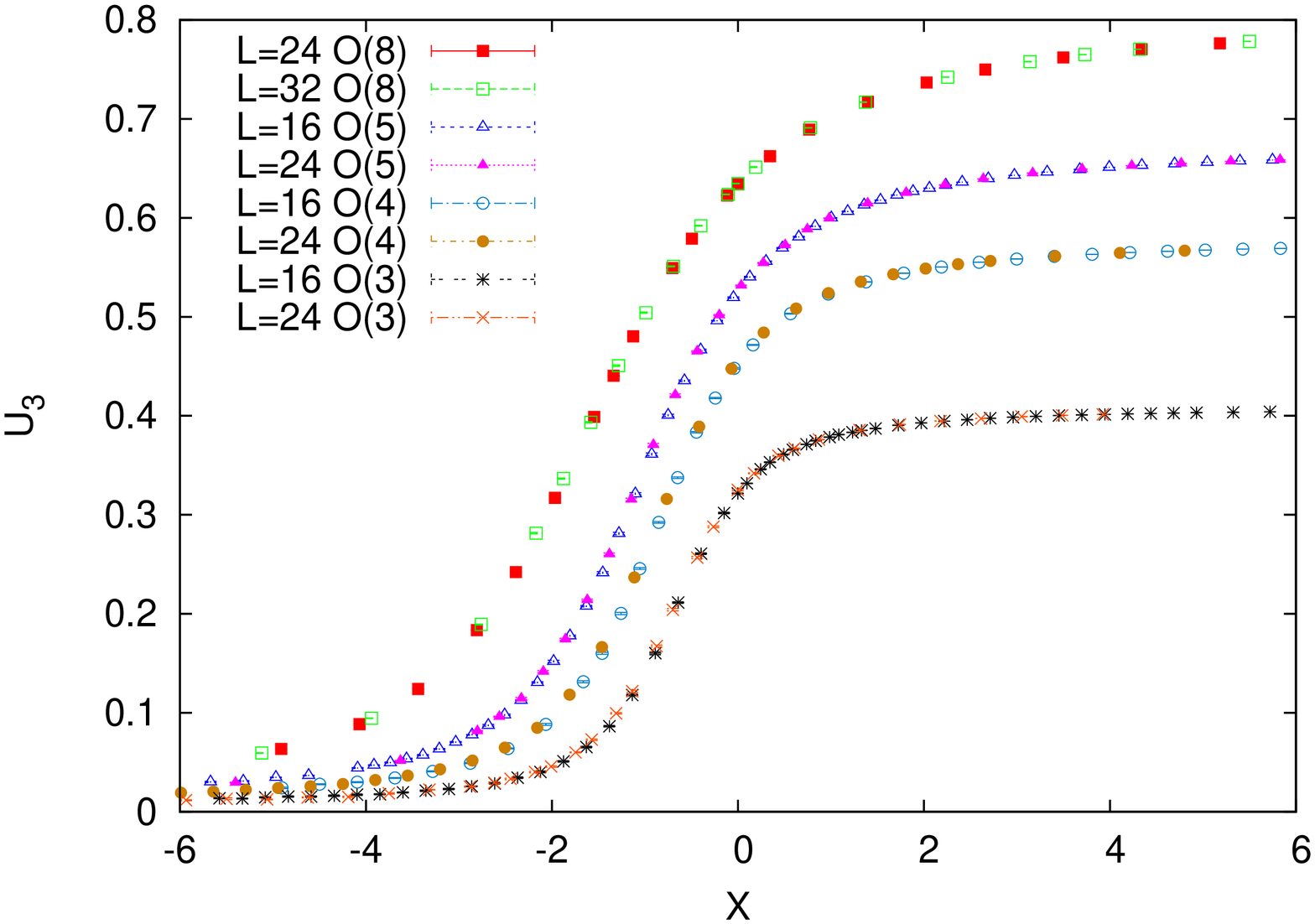} \\
\includegraphics*[scale=\graphicscale,angle=-90]{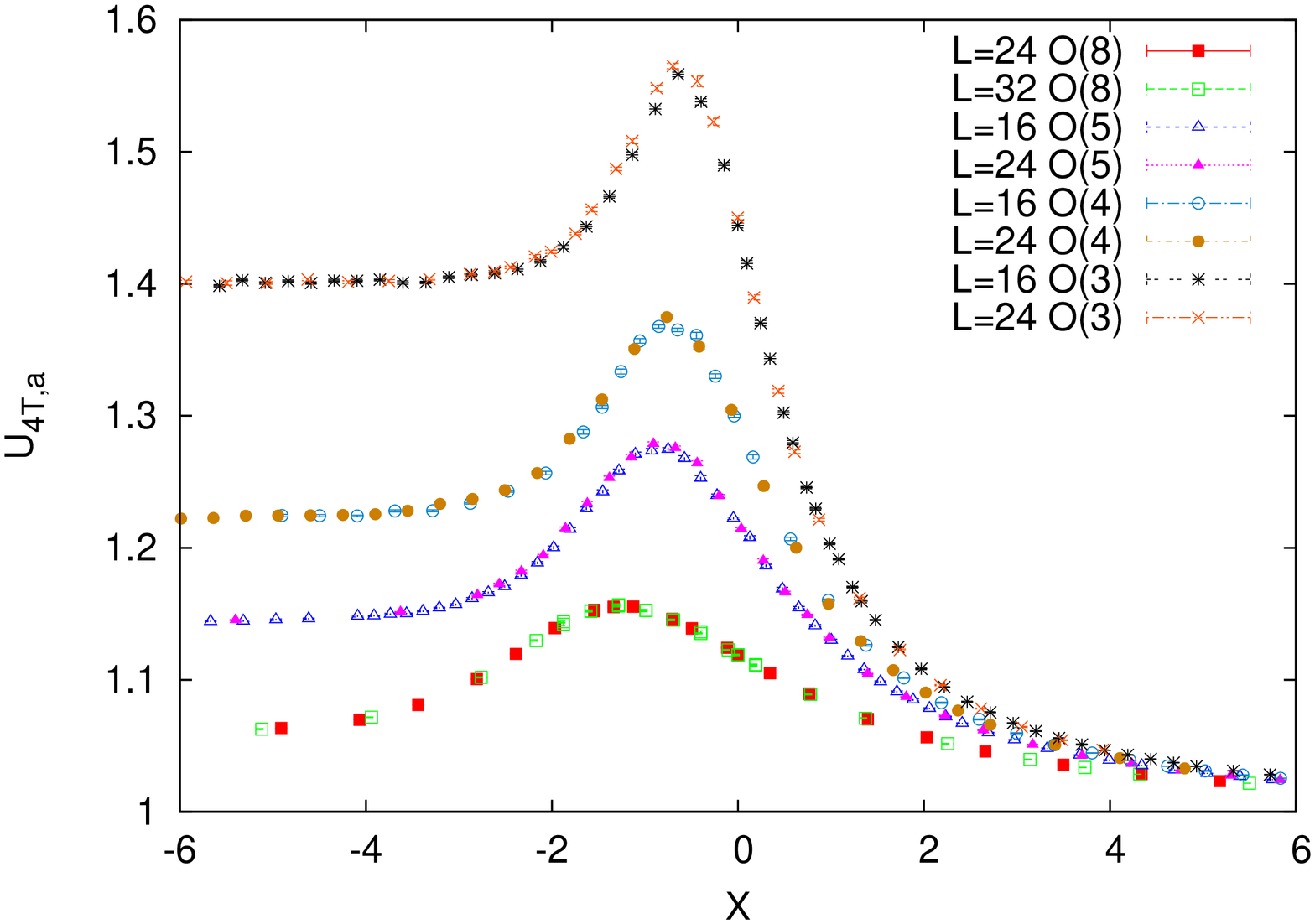} &
\includegraphics*[scale=\graphicscale,angle=-90]{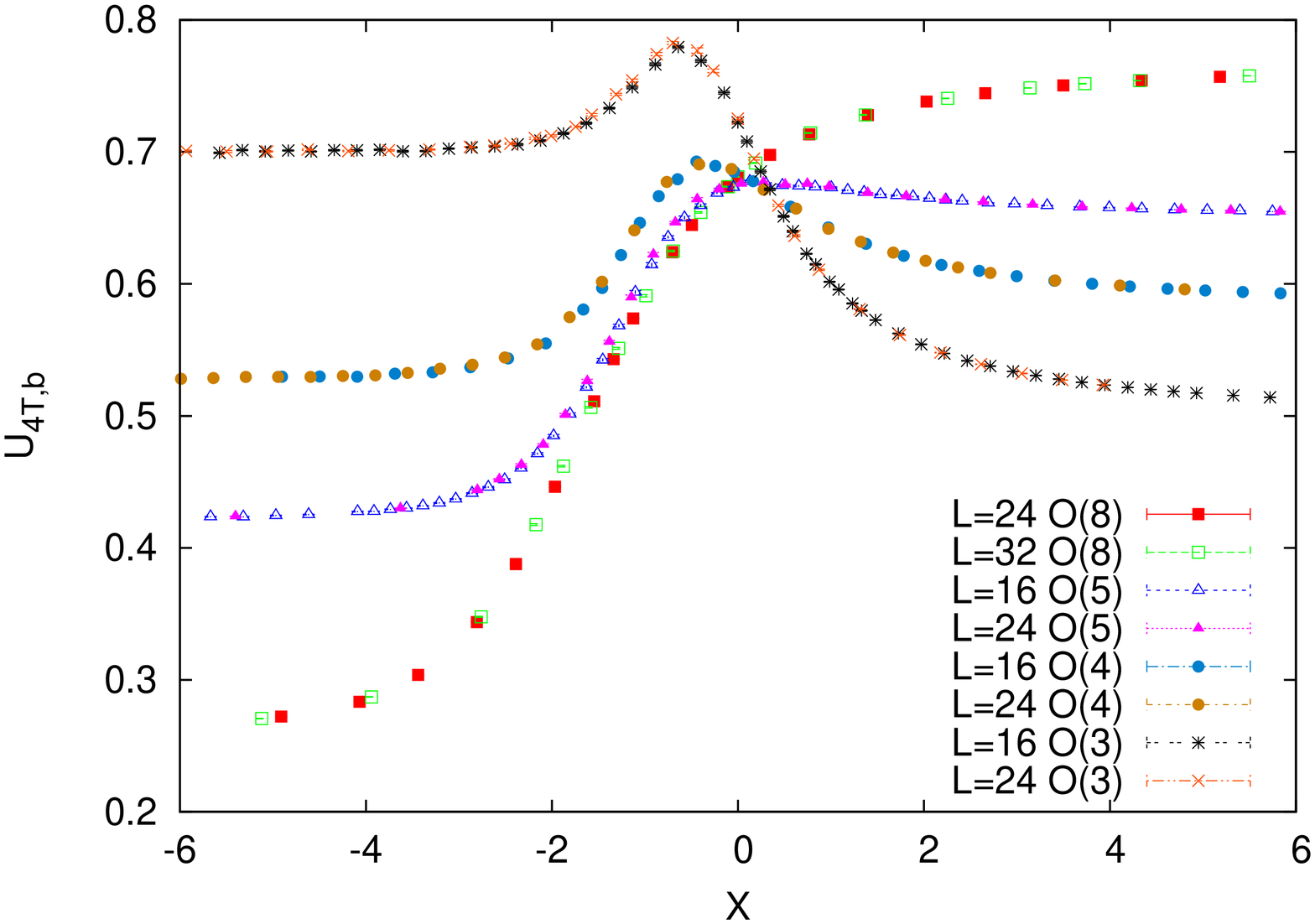} \\
\end{tabular}
\caption{
Plot of $\xi_T/L$ (top left), 
$U_{3T}$ (top right), 
$U_{4T,a}$ (bottom left), and
$U_{4T,b}$ (top right) 
as a function of $X = (\beta - \beta_c) L^{1/\nu}$ 
for $N=3,4,5$ and 8.}
\label{VarTvsX}
\end{figure*}

We have computed several scaling functions for $N=3,4,5,8$ 
on cubic lattices of size $L$, with 
$L$ in the interval $16\le L \le 32$. We use periodic boundary conditions.
In Figs.~\ref{RatioxivsX-suppl} and 
\ref{VarTvsX} we report the scaling functions as a function of $X$. We have
used the following values for $\beta_c$ and $\nu$:
$\beta_c = 0.69302(3)$  \cite{BFMM-96-suppl,BC-97-suppl}
and $\nu = 0.7112(5)$ \cite{CHPRV-02-suppl} for $N=3$;
$\beta_c = 0.93586(1)$ \cite{BFMM-96-suppl,CPRV-96-suppl}
and $\nu = 0.749(2)$  \cite{Hasenbusch-01-suppl}
for $N = 4$;
$\beta_c = 1.18138(3)$ and $\nu = 0.779(3)$ \cite{HPV-05-suppl}  
for $N = 5$;
$\beta_c = 1.92677(2)$ and $\nu = 0.85(2) $ \cite{DPV-15-suppl}
for $N=8$.
Note that, on the scale of the figure, differences on the value of $\nu$
of 1\% cannot be distinguished. 

\begin{table}
\caption{Estimates of several renormalization-group invariant quantities at the 
critical point $X = 0$. }
\label{tableRstar-suppl}
\begin{tabular}{ccccc}
\hline\hline
       & $N=3$              &   $N=4$  &$N=5$ &  $N=8$   \\
\hline
$U_{4T,a}^*$     & 1.458(12)&  1.287(7)& 1.218(2) & 1.117(2) \\
$U_{4T,b}^*$     & 0.729(6) &  0.684(2)& 0.677(3) & 0.681(2) \\
$U_{3T}^*$       & 0.334(10)&  0.462(7)& 0.529(3) & 0.636(2) \\
$(\xi_T/L)^*$    & 0.213(5) &  0.221(4)& 0.222(1) & 0.236(1) \\
$(\xi_V/\xi_T)^*$& 2.68(3)  &  2.50(2) & 2.41(1)  & 2.21(1)   \\
\hline\hline
\end{tabular}
\end{table}

In spite of the fact that lattices are relatively small, we observe a 
very good scaling. We have also determined the value of the different quantities
at the critical point, see Table~\ref{tableRstar-suppl}. The results have been 
obtained by extrapolating the finite-$L$ data assuming that scaling corrections
behave as $L^{-\omega}$. The quoted error includes the statistical error,
the interpolation error of the data, the error on $\beta_c$ and $\nu$, and 
the extrapolation error. The latter has been conservatively estimated as 
the difference between the extrapolated value and the value that the observable
takes on the largest lattice.

\begin{figure*}[!tbp]
\begin{tabular}{cc} 
\includegraphics*[scale=\graphicscale,angle=-90]{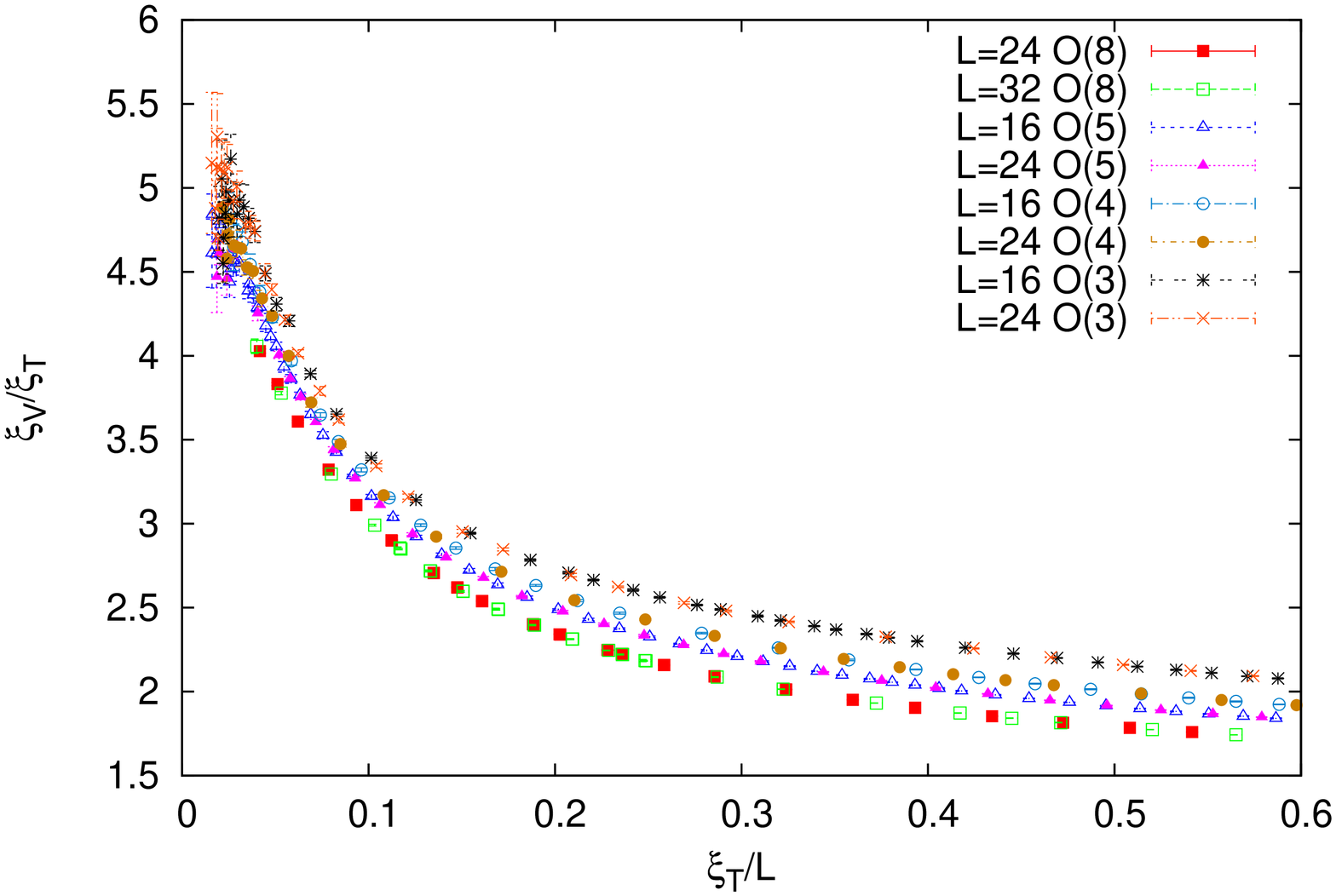} &
\includegraphics*[scale=\graphicscale,angle=-90]{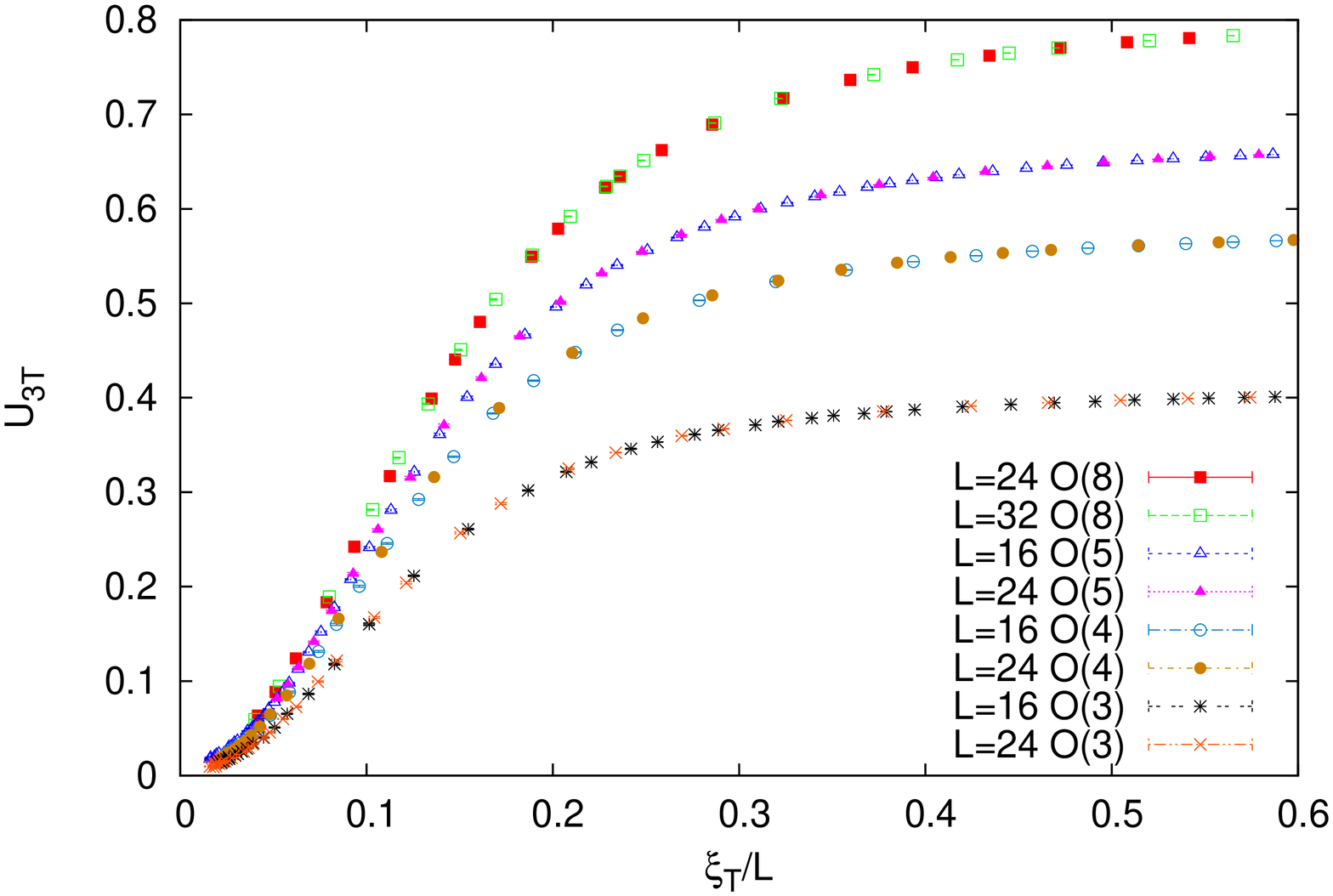} \\
\includegraphics*[scale=\graphicscale,angle=-90]{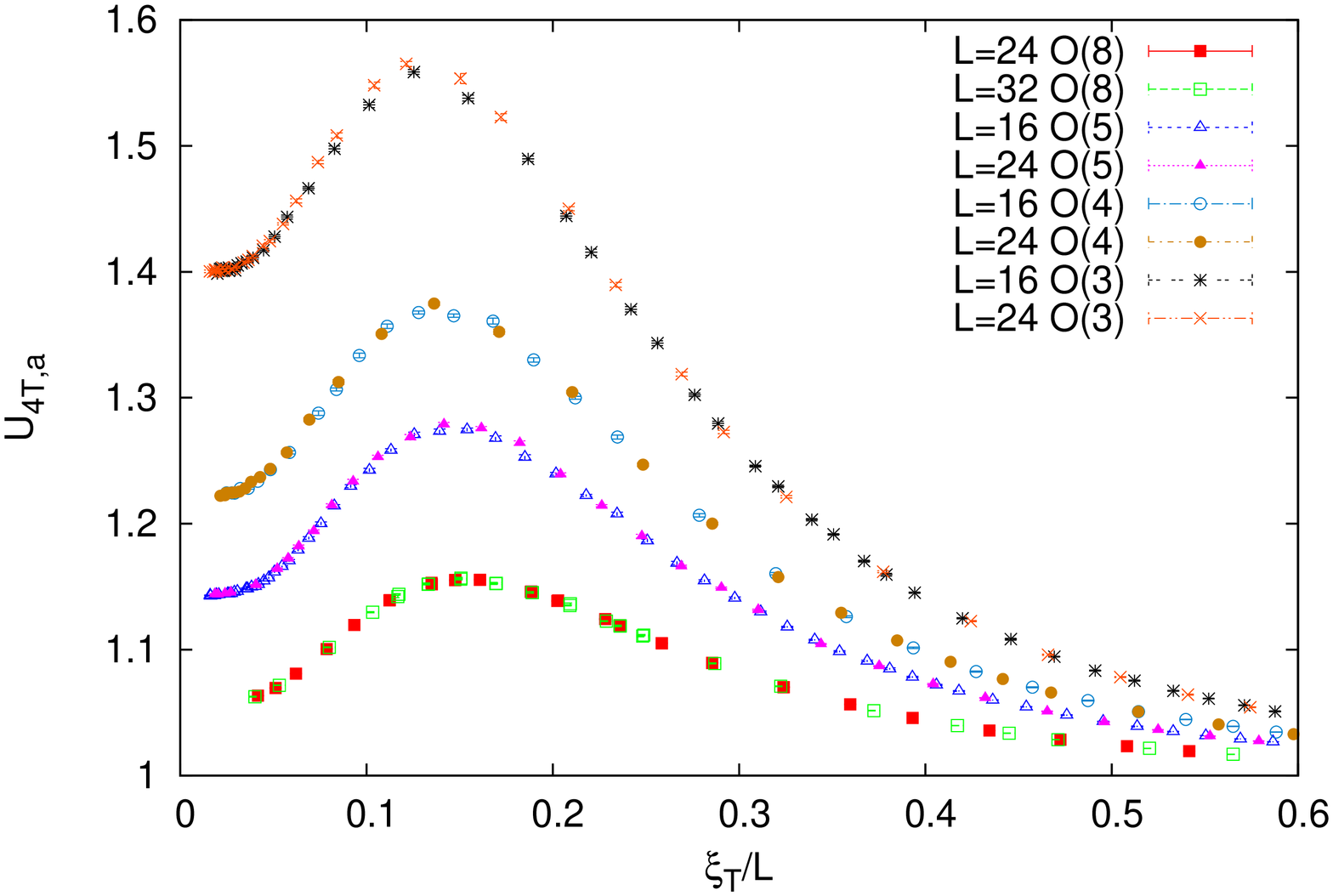} &
\includegraphics*[scale=\graphicscale,angle=-90]{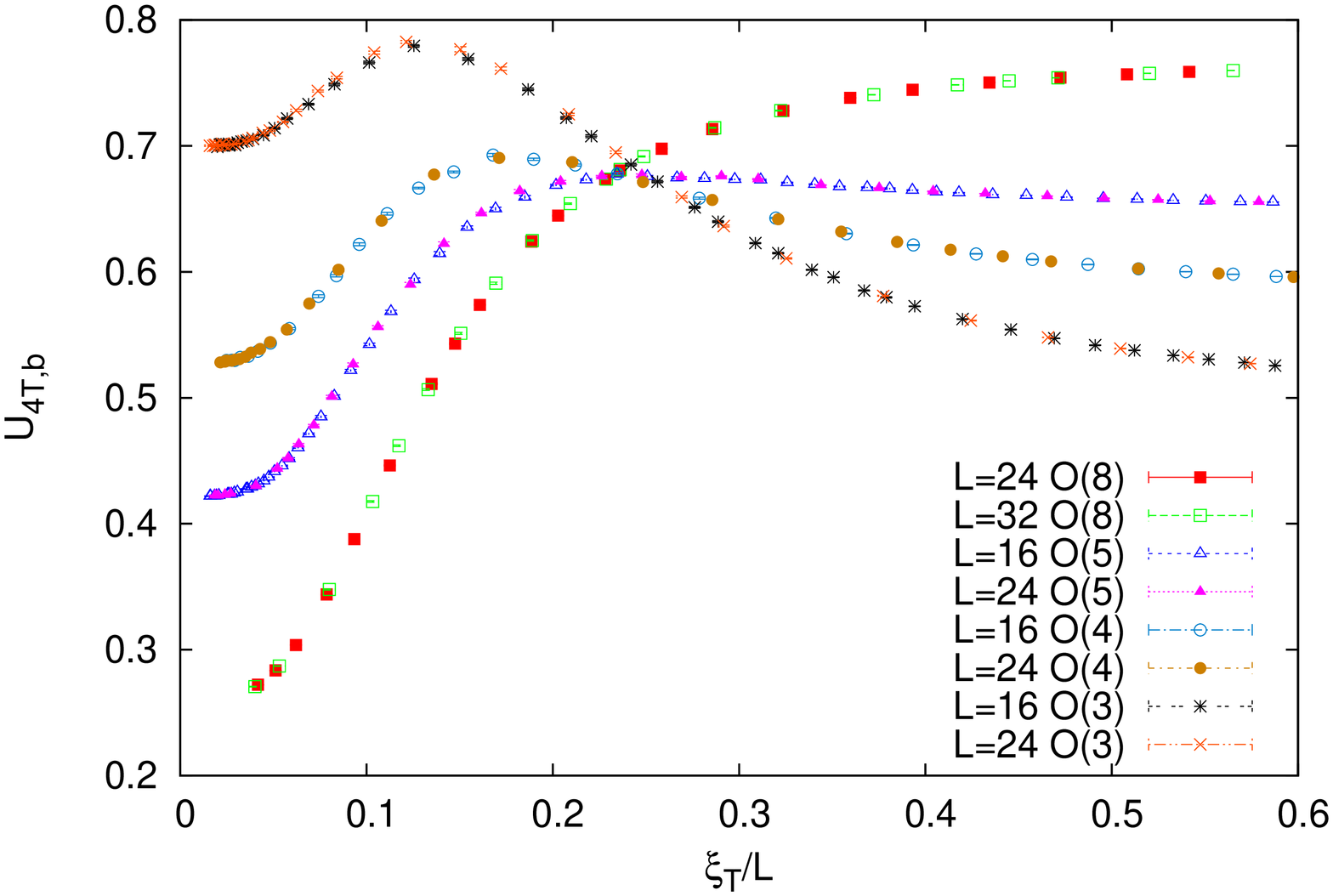} \\
\end{tabular}
\caption{
Plot of $\xi_V/\xi_T$ (top left), 
$U_{3T}$ (top right), 
$U_{4T,a}$ (bottom left), and
$U_{4T,b}$ (top right) 
as a function of $\xi_T/L$
for $N=3,4,5$ and 8.}
\label{VarTvsRxiT-suppl}
\end{figure*}

In Fig.~\ref{VarTvsRxiT-suppl} we report the same invariant ratios
as a function of $\xi_T/L$. Some numerical values (extrapolations and errors
have been computed as before) are reported in Table 
\ref{table:scalingfun-suppl}.

\begin{table*}
\label{table:scalingfun-suppl}
\caption
{We report some numerical values for the scaling function $g_R(x)$ 
($x = \xi_T/L$) defined in Eq.~(\ref{R-vsxiTsuL-suppl}) for $N=3,4,5,8$. 
If the function has a maximum, in the second column we report the 
position $x_{\rm max}$ of the maximum and $\hbox{Max }=g_R(x_{\rm max})$.
In the last five columns we report $g_R(x)$ for $x = 0.1,0.2,0.3,0.4.,0.5$.}
\begin{tabular}{cccccccc}
\hline\hline
 $N$ & $x_{\rm max}$ & Max & $x = 0.1$  &
  $x = 0.2$  &
  $x = 0.3$  &
  $x = 0.4$  &
  $x = 0.5$  \\
\hline
\multicolumn{8}{c}{$U_{3T}$} \\
  3 & & & 0.162(4)  & 0.325(8) & 0.372(2) & 0.390(2) & 0.398(1) \\
  4 & & & 0.212(1)  & 0.422(8) & 0.517(2) & 0.548(2) & 0.561(1) \\
  5 & & & 0.246(7)  & 0.502(7) & 0.597(3) & 0.634(2) & 0.654(1) \\
  8 & & & 0.268(1)  & 0.576(3) & 0.703(2) & 0.753(1) & 0.775(1) \\
\hline
\multicolumn{8}{c}{$U_{4T,a}$} \\
  3 & 0.132(2)  & 1.60(3)  & 1.57(3)   & 1.49(2)   & 1.261(2) & 1.146(5) & 
                 1.080(1) \\
  4 & 0.141(2)  & 1.39(2)  & 1.336(5)  & 1.317(2)  & 1.178(4) & 1.099(2) &
                 1.054(1) \\
  5 & 0.155(1)  & 1.29(1)  & 1.250(7)  & 1.256(2)  & 1.144(4) & 1.075(2) &
                 1.041(1) \\
  8 & 0.154(1) & 1.17(1) & 1.1269(3) & 1.142(2) & 1.089(3) & 1.044(1) &
                 1.0239(4) \\
\hline
\multicolumn{8}{c}{$U_{4T,b}$} \\
  3 & 0.132(2) & 0.80(2)   & 0.78(2)   & 0.745(1)  & 0.631(1) & 0.573(2) &
                 0.540(1) \\
  4 & 0.205(1) & 0.692(6)  & 0.626(3)  & 0.685(3)  & 0.649(2) & 0.622(2)&  
                 0.604(1)  \\
  5 & 0.25(4) &  0.68(1)   & 0.549(7)  & 0.679(9)  & 0.678(4) & 0.665(2)&  
                 0.658(1)  \\
  8 & & & 0.407(1) & 0.644(3) & 0.721(2) & 0.746(1)&  0.7562(2) \\
\hline
\multicolumn{8}{c}{$\xi_V/\xi_T$} \\
  3 & & & 3.37(5) & 2.71(2) & 2.46(1)) & 2.296(8) &  2.171(8) \\
  4 & & & 3.23(5) & 2.62(3) & 2.30(1)  & 2.125(4) &  2.008(7) \\
  5 & & & 3.18(2) & 2.49(2) & 2.20(1)  & 2.029(3) &  1.918(6) \\
  8 & & & 3.01(2) & 2.34(1) & 2.05(1)  & 1.893(3) &  1.788(4) \\
\hline\hline
\end{tabular}
\end{table*}

\subsection{CP$^{N-1}$ Binder parameters in the O($2N$) theory} 

In the limit $\beta_g\to\infty$, the gauge field $\lambda_{{\bm x},\mu}$ 
can be set 
equal to one and the model reduces to a vector model with a $2N$ dimensional 
real spin field. We wish now to compute the limiting behavior 
of the scaling functions for the CP$^{N-1}$  Binder parameters. As we have done 
for the $N$-vector theory, we define three Binder parameters. If 
\begin{equation}
Q^{ab}_{\bm x} = \bar{z}^a_{\bm x} \bar{z}^b_{\bm x} - 
    {1\over N} \delta^{ab}  \qquad 
   \Theta_{CP}^{ab} = \sum_{\bm x} Q^{ab}_{\bm x} ,
\end{equation}
we define
\begin{equation}
U_{3CP} = {\langle \hbox{Tr } \Theta^3_{CP} \rangle \over 
    \langle \hbox{Tr } \Theta^2_{CP}\rangle^{3/2}},
\end{equation}
\begin{equation}
U_{4CP,a} = {\langle [\hbox{Tr } \Theta^2_{CP}]^2\rangle  \over 
      \langle \hbox{Tr }\Theta^2_{CP}\rangle^2 } \qquad
U_{4CP,b} = {\langle \hbox{Tr } \Theta^4_{CP} \rangle \over 
  \langle \hbox{Tr } \Theta^2_{CP} \rangle^2}. \qquad
\end{equation}
The quantity named here $U_{4CP,a}$ is the Binder parameter $U_4$ presented in 
the main text.  For $N=2$, we have $U_{3CP} = 0$ and $U_{4CP,a} = 2 U_{4CP,b}$.

In the high-temperature phase we have $U_{3CP} = 0$, 
\begin{equation}
U_{4CP,a} = {N^2 + 1 \over N^2 - 1} \quad
U_{4CP,b} = {2 N^2 - 3 \over N(N^2 - 1)}.
\end{equation}
In the low-temperature phase we have $U_{4CP,a} = 1$,
\begin{eqnarray}
U_{3CP} &=& {N-2\over \sqrt{N(N-1)}} , \\
U_{4CP,b} &=& {N^2 - 3 N + 3\over N (N-1)}.
\end{eqnarray}
To compute the scaling functions associated with
 these quantities in the O($2N$) theory,
we use the mapping 
\begin{equation}
z_{\bm x}^a = s^a_{\bm x} + i s^{a+N}_{\bm x},
\end{equation}
where $1\le a \le N$ and $s^{\alpha}_{\bm x}$ is a $2N$-dimensional real unit 
vector.  It is then easy to verify that
\begin{equation}
\Theta_{CP}^{ab} = \Theta^{ab} + \Theta^{a+N,b+N} + i \Theta^{a,b+N} 
   - i \Theta^{a+N,b},
\end{equation}
where $\Theta^{ab}$ is the $2N$-dimensional real quantity defined in
Eq.~(\ref{defTheta-suppl}). Squaring the previous relation and taking the 
trace, we obtain 
\begin{eqnarray}
\hbox{Tr } \Theta_{CP}^2 &=& \hbox{Tr } \Theta^2 + 
\label{Eq-TRCP-TRO2N-suppl} \\
&& + 
\sum_{a,b=1}^N (\Theta^{ab} \Theta^{a+N,b+N} -
             \Theta^{a,b+N} \Theta^{a+N,b}).
\nonumber
\end{eqnarray}
In this relation, the trace in the l.h.s. is performed in the 
CP$^{N-1}$ theory (indices go from 1 to $N$), while the trace in the 
r.h.s. is performed in the O($2N$) theory (indices go from 1 to $2N$). 
The presence of the additional terms in Eq.~(\ref{Eq-TRCP-TRO2N-suppl}) 
explains why the relation between the CP$^{N-1}$ Binder parameters and the 
tensor Binder parameters in the O($2N$) theory is not trivial.

\begin{figure*}[!tbp]
\begin{tabular}{cc} 
\includegraphics*[scale=\graphicscale,angle=-90]{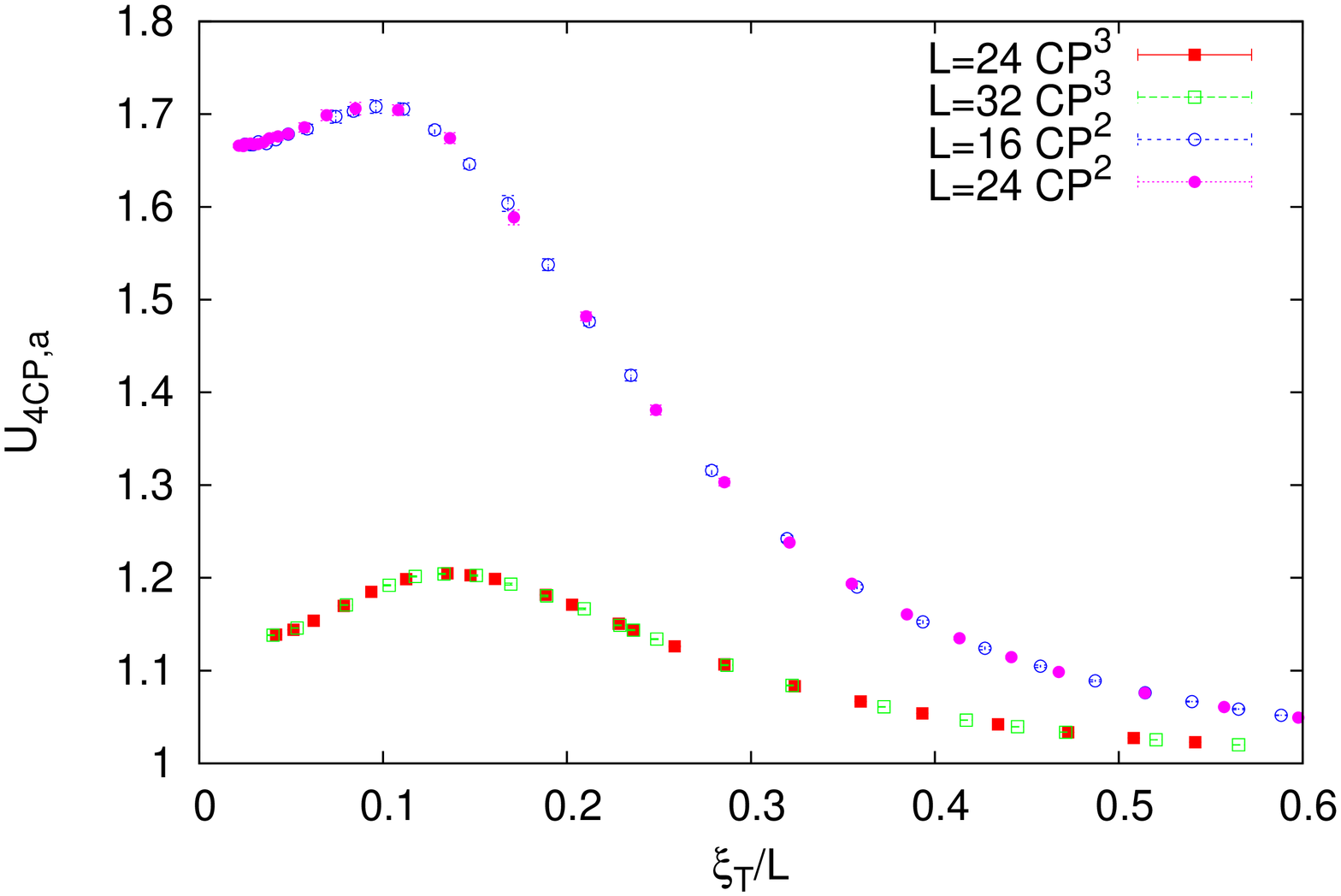} &
\includegraphics*[scale=\graphicscale,angle=-90]{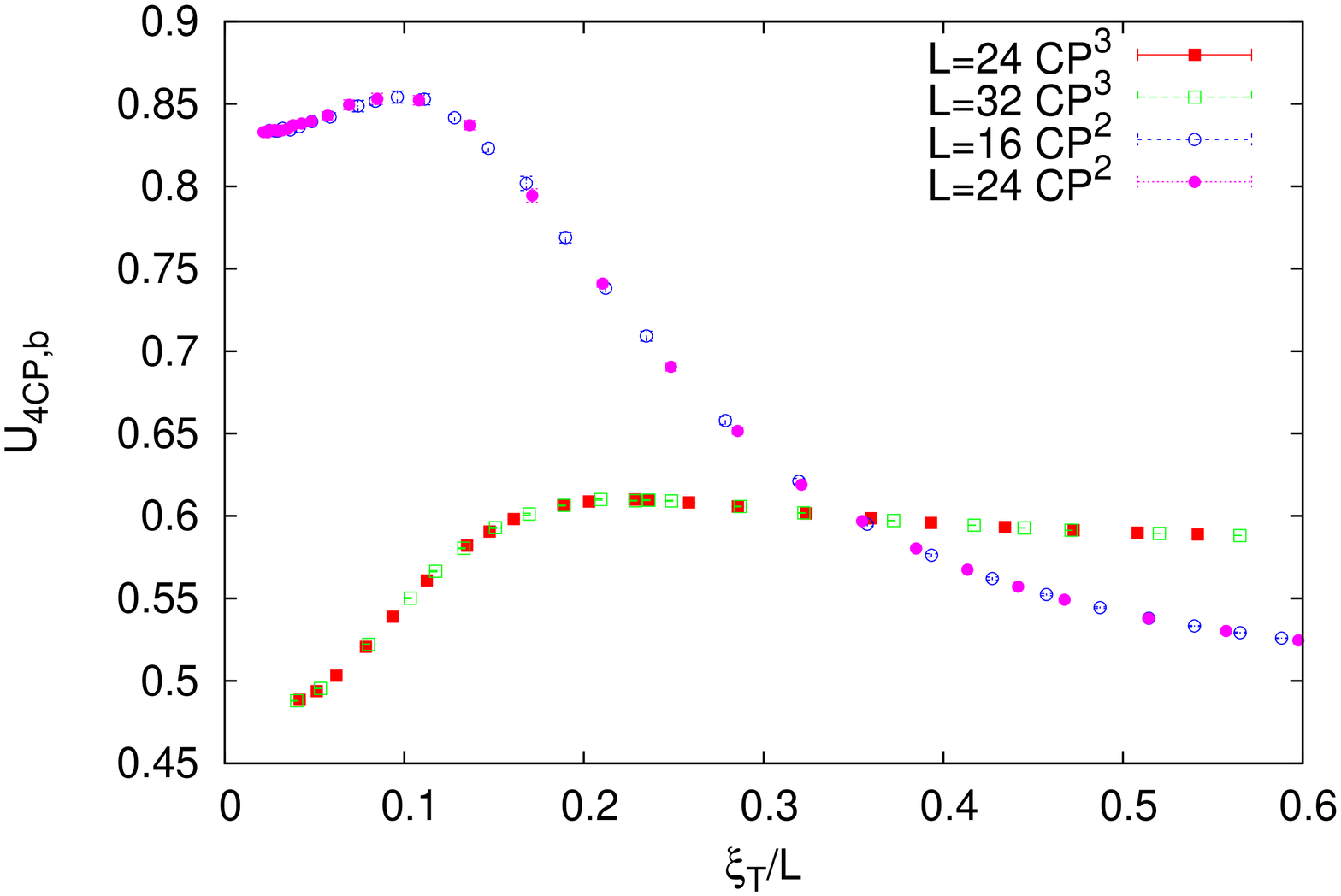} \\
\end{tabular}
\caption{Scaling functions of the CP$^{N-1}$ Binder parameters in the 
O($2N$) model, computed using O($2N$) tensor data and 
Eq.~(\protect\ref{relO2n-CPN-suppl}). For $N=2$ we have 
$U_{4CP,b} = U_{4CP,a}/2$. For $\xi_T/L\to 0$, we have 
$U_{4CP,a} \approx 1.667$, $U_{4CP,b} \approx 0.833$ for $N=2$ and 
$U_{4CP,a} \approx 1.133$, $U_{4CP,b} \approx 0.483$ for $N=4$. In 
the limit $\xi_T/L\to \infty$, $U_{4CP,a}=1$ and 
$U_{4CP,a}=0.5, 0.583$ for $N=2$, 4, respectively.}
\label{CPNscaling-suppl}
\end{figure*}

Using the O($2N$) invariance of the model  we obtain the relations
\begin{eqnarray}
&& \langle \hbox{Tr } \Theta_{CP}^2 \rangle = {2 (N-1) \over 2 N - 1}
   \langle \hbox{Tr } \Theta^2 \rangle, \nonumber \\
&& \langle \hbox{Tr } \Theta_{CP}^3 \rangle = 
 {2 (N-2)\over 2 N - 1} \langle \hbox{Tr } \Theta^3 \rangle, \nonumber \\
&& \langle [\hbox{Tr } \Theta_{CP}^2]^2 \rangle = 
   {4 N\over (2 N - 3) (4 N^2 - 1)} \times 
\nonumber \\
 && \quad \times 
   \left[(2 N^2 - 5 N + 4) \langle [\hbox{Tr } \Theta^2]^2 \rangle -
        2 \langle \hbox{Tr } \Theta^4  \rangle \right], \nonumber \\
&& \langle \hbox{Tr } \Theta_{CP}^4 \rangle = {2\over (2 N - 3) (4 N^2 - 1)}
\times \nonumber \\
&& \quad \times 
\left[(4 N^2 - 3 N -6) \langle [\hbox{Tr } \Theta^2]^2 \rangle + 
\right. \\
&& \qquad \qquad \left. + 
        2 (2 N^3 - 9 N^2 + 6 N + 6) 
  \langle \hbox{Tr } \Theta^4 \rangle \right]. \nonumber 
\end{eqnarray}
We obtain therefore the relations, which are exact for the O($2N$) theory,
between CP$^{N-1}$ and tensor O($2N$) Binder parameters:
\begin{eqnarray}
&& U_{3CP} = {N-2\over N-1} \sqrt{2 N - 1\over 2 (N-1)} \, U_{3T}, 
\nonumber \\
&& U_{4CP,a} = {N (2 N - 1) \over (N-1)^2 (2 N + 1) (2 N - 3)} \times
\nonumber \\
&& \qquad \times
 \left[(2 N^2 - 5 N + 4) U_{4T,a}  -
        2 U_{4T,b} \right], \nonumber \\
&& U_{4b,CP} =
  {2 N - 1 \over 2 (N-1)^2 (2 N + 1) (2 N - 3)} \times \nonumber \\
&& \qquad \times
 \left[(4 N^2 - 3 N -6) U_{4T,a} + \right. \nonumber \\
&& \qquad \quad + \left.
        2 (2 N^3 - 9 N^2 + 6 N + 6) U_{4T,b} \right].
\label{relO2n-CPN-suppl}
\end{eqnarray}
For the CP$^1$ model ($N=2$) the only independent Binder parameter
$U_{4CP,a}$ is related to the tensor Binder parameters in the O(4) theory
by
\begin{equation}
U_{4CP,a} = {12\over 5} (U_{4T,a} - U_{4T,b}).
\end{equation}
For the CP$^3$ model ($N=4$) the relation between CP$^3$ and O(8) 
Binder parameters is
\begin{eqnarray}
U_{3CP} &=& \sqrt{14\over 27}\, U_{3T}, \nonumber \\
U_{4CP,a} &=& {448\over 405} U_{4T,a} - {56\over 405} U_{4T,b}, \nonumber \\
U_{4CP,b} &=& {161\over 405} U_{4T,a} + {98\over 405} U_{4T,b}.
\end{eqnarray}
For the comparison it is also important to relate the correlation length. 
In the CP$^{N-1}$ it is defined from the correlation function of $Q_x$:
\begin{equation}
G_{CP}({\bm x}) = \langle \hbox{Tr } Q_{\bm 0}\ Q_{\bm x}\rangle.
\end{equation}
In the O($2N$) theory, we have the relation
\begin{equation}
G_{CP}({\bm x}) = {2 (N-1)\over 2N-1} G_T({\bm x}),
\end{equation}
As the correlation length does not depend on the normalization of the 
correlation function, the correlation length $\xi_{CP}$ computed from 
$G_{CP}({\bm x})$ is identical to the O($2N$) tensor correlation length $\xi_T$.
The scaling functions for the CP$^{N-1}$ Binder parameters $U_{4CP,a}$ and 
$U_{4CP,b}$ are reported in Fig.~\ref{CPNscaling-suppl} for $N=2$ and 4.

\begin{widetext}

\end{widetext}


\begin{thebibliography}{99}
\bibitem{footnoteH} 
Note that our coupling $\beta_g$ differs by a factor of 2 
from the one used in other works.
If $\lambda_{{\bm x},\mu} = \exp(i \theta_{{\bm x},\mu})$, the gauge part 
of the Hamiltonian reads 
$H_{\rm gauge} = -2 \beta_g \cos \theta_{{\bm x},\mu\nu}$, with 
$\theta_{{\bm x},\mu\nu} = \theta_{{\bm x},\mu} - 
\theta_{{\bm x}+\hat{\nu},\mu}  - \theta_{{\bm x},\nu} + 
\theta_{{\bm x}+\hat{\mu},\nu}$.
\bibitem{footnoteold} In Ref.~\cite{KHI-11,ODHIM-07,TIM-05}, the
  system sizes were quite small and the data had a limited precision,
  preventing the authors to obtain a satisfactory understanding of the nature
  of the transitions and of the crossover behaviors along the
  transition lines.
\end{thebibliography}
\end{document}